\documentclass[preprint,1p,times,twocolumn]{lalaarticle}
\usepackage[applemac]{inputenc}
\usepackage{syntax}
\usepackage[cyr]{aeguill}
\usepackage{listings}
\usepackage{multirow}
\usepackage{amssymb}
\usepackage{url}

\lstdefinelanguage{smtlib} {morekeywords={declare,datatypes,sort,store,select,fun,const,Int,Bool,assert,forall,exists,and,or,not,check,sat,get,model,pattern,array,=,=>,>,<},
sensitive=false,
morecomment=[l][\small\itshape]{;}}
\usepackage{lscape}

\usepackage[cyr]{aeguill}
\usepackage{microtype}

\lstdefinelanguage{Alloy}
{morekeywords={abstract, all, and, as, assert, but, check, disj, else,
  exactly, extends, fact, for, fun, iden, if, iff, implies, in, Int,
  int, let, lone, module, no, none, not, one, open, or, part, pred,
  run, seq, set, sig, some, sum, then, univ},
sensitive=true,
morecomment=[l][\small\itshape]{--},
morecomment=[l][\small\itshape]{//},
morecomment=[s][\small\itshape]{/*}{*/},
tabsize=2,
columns=fullflexible,
literate={->}{{$\to$}}1 {^}{{$\mspace{-3mu}\widehat{\quad}\mspace{-3mu}$}}1
 {<}{$<$ }2 {>}{$>$ }2 {>=}{$\geq$ }2 {=<}{$\leq$ }2
 {<:}{{$<\mspace{-3mu}:$}}2 {:>}{{$:\mspace{-3mu}>$}}2
 {<=>}{{$\Leftrightarrow$ }}2 {=>}{{$\Rightarrow$ }}2 {+}{$+$ }2 {++}{{$+\mspace{-8mu}+$ }}2
 {\~}{{$\mspace{-3mu}\widetilde{\quad}\mspace{-3mu}$}}1
 {!=}{$\neq$ }2 {*}{${}^{\ast}$}1 {.}{$\cdot$}1
 {\#}{$\#$}1
}
\usepackage{lscape}

\usepackage{amssymb}

\journal{Science of Computer Programming}

\begin{document}

\begin{frontmatter}

\title{A Direct Symbolic Execution of SQL Code \\ for Testing of Data-Oriented Applications}
\tnotetext[label1]{}
\author{Michaël Marcozzi\fnref{label2}}
\ead{michael.marcozzi@unamur.be}
\fntext[label2]{F.R.S.-FNRS Research Fellow}
\author{Wim Vanhoof}
\ead{wim.vanhoof@unamur.be}
\author{Jean-Luc Hainaut}
\ead{jean-luc.hainaut@unamur.be}
\address{Faculty of Computer Science\\
University of Namur\\
Rue Grandgagnage, 21\\
5000 Namur, Belgium}

\begin{abstract}
Symbolic execution is a technique which enables automatically generating test inputs (and outputs) exercising a set of execution paths within a program to be tested. If the paths cover a sufficient part of the code under test, the test data offer a representative view of the program's actual behaviour, which notably enables detecting errors and correcting faults. Relational databases are ubiquitous in software, but symbolic execution of pieces of code that manipulate them remains a non-trivial problem, particularly because of the complex structure of such databases and the complex behaviour of SQL statements. In this work, we define a direct symbolic execution for database manipulation code and integrate it with a more traditional symbolic execution of normal program code. The database tables are represented by relational symbols and the SQL statements by relational constraints over these symbols and the symbols representing the normal variables of the program. An algorithm based on these principles is presented for the symbolic execution of Java methods that implement business use cases by reading and writing in a relational database, the latter subject to data integrity constraints. The algorithm is integrated in a test generation tool and experimented over sample code. The target language for the constraints produced by the tool is the SMT-Lib standard and the used solver is Microsoft Z3. The results show that the proposed approach enables generating meaningful test data, including valid database content, in reasonable time. In particular, the Z3 solver is shown to be more scalable than the Alloy solver, used in our previous work, for solving relational constraints. 
\end{abstract}


\end{frontmatter}

‡
\newpage
\section{Introduction}
\label{Introduction}
In current software development practice, testing~\cite{Jorgensen2008,testing-book-kaner} remains the primary approach to detect errors in software. Testing being a complex and costly activity has motivated much research on efficient techniques to automate all aspects of the software testing process \cite{Ramler:2006}. 
Of particular interest is the automation of \emph{test data generation} \cite{adequacy-zhu} for functional testing of units of code, where the idea is to automatically generate a representative set of inputs (and outputs) for a unitary program fragment under test (typically a function or method). These data can subsequently be compared with the function's expected behaviour in order to detect errors and correct faults. Moreover, once a suitable set of test data has been generated (and verified), it can be used as reference data for continued (regression) testing of the code.

While different approaches exist towards the automatic generation of test data, \emph{symbolic execution}~\cite{King-1976} has been recognised as a promising technique for so-called \emph{white-box} or \emph{glass-box} testing~\cite{symbex-review1,symbex-review2,symbex-review3,se-synt,sybolic-execution}, where the idea is to generate test data that in some way \emph{cover} a sufficiently large part of the control-flow graph of the function under test~\cite{adequacy-zhu}. 
The symbolic execution process traverses the control-flow graph of a program or function by executing the code over symbolic values instead of concrete values~\cite{King-1976}. Each time a control dependency is encountered, the symbolic execution process proceeds along one of the possible paths, thereby  generating constraints upon the symbolic values such that when the program variables have values that satisfy these constraints, the real execution would proceed along the selected path. For each path, the constraints that have been collected along that path are regrouped in a so-called \emph{path-constraint} which is subsequently solved, resulting in a set of concrete test values for the program variables that make the real execution proceed along the given path. The process terminates when test data have been generated for a sufficiently large (and diverse) number of paths through the control-flow graph, according to some coverage criterion~\cite{adequacy-zhu}. 
Test data generation based on symbolic execution is now at the core of various popular open-source and commercial testing tools, some being used in an industrial context, notably at Microsoft and NASA \cite{se-synt}. \\
‡
 
Symbolic execution of imperative code has been widely studied \cite{cbt1,yices,Cadar06exe,heuristicPaths2}, as well as generalised to lower-level or higher-level programming paradigms, such as for x86 assembly \cite{sage} or Java \cite{cbt4,Java-tool2B} and C\# \cite{cbt2} object-oriented code. In so-called database programs, such classical code is mixed with SQL code to interact with a relational database. Enabling symbolic execution in the presence of SQL statements among the classical statements is a non-trivial extension of existing symbolic execution techniques. 
 
From a theoretical standpoint, deciding the satisfiability of an SQL query is not computationally possible in general \cite{dbs}, so that generating test data for any particular execution path, in the presence of SQL statements in the path, is not generally computable either \cite{qex}. From a more practical standpoint, the database can be seen as nothing else than a particular kind of container for some of the values manipulated by the program. During symbolic execution, these values must thus be represented symbolically and subsequently constrained to enable the proper generation of test data, including an input and output database content. The symbolic representation of these values raises difficulties as the database is a container of particularly complex shape: its content must obey the so-called \emph{database schema}, defined using SQL DDL code \cite{Date:2003}. This database schema defines a set of tables where the database content will be stored. Each table can be seen as a container for a mathematical relation, i.e., a set of tuples with no limit on the potential number of tuples. The schema typically also describes a set of \emph{data integrity constraints} that must be enforced by the relations in the tables, like the primary key, foreign key or check constraints. These constraints are particularly complex as they are quantified first-order logic constraints. For instance, a primary key constraint states that for all couples of tuples in the relation represented by a table, the value of the primary key fields cannot be all equal. 

Moreover, a program typically interacts with the database by using SQL \cite{Date:2003} \emph{query statements} and \emph{DML statements} that are embedded in the program's source code. As such, SELECT queries enable gathering values from the database tables in order to copy them into the program variables. DML statements INSERT, UPDATE and DELETE enable modifying the content of the database tables, typically a function of the value of the program variables. Symbolic execution of such SQL statements raises practical difficulties as well, due to their complex behaviour. Firstly, SQL is a declarative language: 
SQL statements express the desired action over the content of the database relations, but they do not make explicit the (often complex) control-flow necessary to compute this action.
In practice, during execution, these SQL statements are sent by the database program -- using a dedicated API -- to the DataBase Management System (DBMS) \cite{Date:2003}, an external component responsible for the interpretation and execution of SQL code over the database. The DBMS keeps an optimised and persistent internal representation of the database and manages concurrent distant accesses by enabling the database programs to use \emph{transactions} \cite{Date:2003}. Secondly, the execution of DML INSERT, UPDATE or DELETE statements by the DBMS does not only consist in modifying the content of the database, but also in checking that these modifications do not leave the database in a state where the integrity constraints defined in the schema are violated. If an integrity constraint violation is detected, the DML statement execution fails and the database remains unmodified. As a consequence, during symbolic execution each INSERT, UPDATE or DELETE statement will have to be treated as an if-then-else statement with a particularly complex condition:

\lstset{language=Java,frame=single,columns=flexible,showstringspaces=false,escapechar=\%}
\begin{lstlisting} 
if (%\textit{Program variables and Database are in a state}%
    %\textit{where the SQL statement will not violate any constraint}%) {
	Execute the SQL statement!
} else {
	Signal a constraint violation!
}
\end{lstlisting} 

Two research approaches have been developed so far to handle the presence of SQL statements within the classical code to be symbolically executed. The first approach \cite{3,xie-vn} introduces new native variables in the classical code to represent the database content and replaces the SQL queries and DML statements by native code acting on these new variables. Classical symbolic execution can then be applied on this \emph{normalised} version of the database program. The second approach \cite{13} considers the result of any query in the tested code as a new \emph{array-like program input}, which can be constrained within the path constraints. Solving any of these path-constraints enables generating input content to insert into the database, so that it will return adequate results for the queries along the considered path.

In this work, we propose a different approach, called \emph{relational symbolic execution}, which enables a direct translation of both SQL queries and DML statements into constraints. Contrary to the normalisation approach, relational symbolic execution makes it possible to test the original code directly, without transforming it. Contrary to the input arrays approach, relational symbolic execution makes it possible to test programs which can also write data into the database, using DML statements.

In practice, relational symbolic execution models every table in the database as a program variable typed as a mathematical relation. Each SQL statement in the program can then be modelled as a relational operation over these \emph{relational variables} as well as the traditional program variables. By defining a symbolic execution over this relational version of the program, we can derive a set of path constraints over the values of both the program's relational and traditional variables. 
The generated path constraints will thus include symbols representing classical program values as well as symbols representing the relations in the database tables.
Furthermore, each path constraint must be combined with schema constraints over the relational symbols, in order to enforce data integrity of the database content. The result is a complex constraint system that mixes traditional constraints over the content of the program's variables with relational constraints over the content of the relational variables.  Each solution, instantiated to the combined constraint system, describes test data, \emph{including} a valid initial and final state for each table in the database, such that when the program is executed with respect to these data values (including the database), the execution will follow the path represented by the constraint system. \\

The main contribution of this work is a relational symbolic execution algorithm, which implements the technique described in the previous paragraph for a precise subset of Java, empowered by the JDBC API for using SQL primitives and transactions. The algorithm has been designed in the context of testing transactional software, where database programs can typically be divided into a set of rather small methods, implementing each a precise business use case, like registering a book borrowing or saving a sale transaction. Such an use case involves a short sequence of SQL actions touching a limited number of entities in the database, and must be executed in real-time. Given the Java/SQL code of a single method implementing a business use case, the SQL DDL code describing the schema of the database manipulated by this method and a finite path in the control flow graph of the method, the algorithm generates the corresponding constraints.

A major challenge for relational symbolic execution is solving the mix of relational and classical constraints produced by the approach. In our previous work \cite{moi-1,moi-scam13}, these constraints were expressed using the Alloy language \cite{alloy-book} and solved using the Alloy analyser \cite{alloy-book}. Recently, it has been proposed \cite{smt-alloy} to translate Alloy constraints into the SMT-Lib standard language \cite{smt2}, as some solvers based on this language, like Microsoft Z3 \cite{z3}, enable detecting the unsatisfiability of the constraints, which is not possible within the Alloy framework. However, \cite{smt-alloy} also advises to use the Alloy analyser for finding solutions to satisfiable constraints, as Z3 had exhibited some limitations for this task, due to the unavoidable presence of quantifiers when translating relational constraints into SMT-Lib. Nevertheless, more recent versions of Z3 take advantage of new model-based quantifier instantiation capabilities (MBQI)  \cite{mbqi}, providing the solver with a promising ability to find efficiently solutions for satisfiable quantified constraints. In this paper, the relational symbolic execution algorithm from our previous work  has been redesigned to generate constraints in the SMT-Lib language and solved using an MBQI-powered version of Z3. Z3 is at the core of several existing state-of-art symbolic execution tools (e.g. \cite{cbt2,sage}). Moreover, it handles a much larger and more various panel of constraints than Alloy, the latter being limited to bounded-scope integer and relational constraints. Using Z3 as a back-end solver does strengthen thus the generalisability of our approach.

A test generation tool based on our new relational symbolic execution algorithm and the Z3 solver has been coded and used to generate test data for a number of sample Java methods and databases, including open-source code extracted from the web. These experiments enabled a comparative evaluation of Alloy and Z3 for solving the generated constraints. They also made it possible to compare, to some extent, the performance of our approach with that of related work based on translation of SQL code into native code. \\

The remainder of this paper is organised as follows. The relational symbolic execution algorithm is described in section \ref{theo}. After discussing the transactional context for which the algorithm was designed, we formally define the part of the Java/SQL syntax that is supported. Then, we systematically describe the constraint generation rules to be used for the symbolic execution of the supported language. The test generation tool based on the algorithm is described, together with our  experimental evaluation of the approach, in section \ref{experimentation}. Finally, some conclusions, related and future work are discussed in section \ref{conclusion}.
\newpage

\section{A relational symbolic execution algorithm for transactional use cases}
\label{theo}

\subsection{Database programs in transactional software}

Interaction between software and a relational database is often said to occur either in an OLTP (On-Line Transaction Processing) or in an OLAP (On-Line Analytical Processing) context. In a nutshell, OLTP corresponds to the case where the database must be read and written to support human activities in real time, like managing concurrent bank transactions, updating customer files or processing ticket registrations. The OLAP context occurs when the data gathered in the database are patiently analysed, using data mining techniques, typically for reporting or strategical purposes, like accounting, budget or marketing. 

Our algorithm has been designed in the context of OLTP. The documentation of Microsoft's DBMS SQL Server 2012\footnote{\url{http://technet.microsoft.com/en-us/library/hh393556(v=sql.110).aspx}} describes database programs designed in an OLTP context as "characterised by small, interactive transactions that generally require sub-second response times". More precisely, a characteristic of the database programs written in an OLTP context is that they can typically be divided into rather small \cite{gray} methods "implementing [business] use cases which execute a sequence of actions whereas each action usually reads or updates only a small set of tuples in the database" \cite{otherApproach}. The purpose of our work is to generate test data for such methods. An example extracted from \cite{otherApproach} of an use case implemented by such a method is a library user that wants to borrow a book. The code checks the user and book data and possibly allows and saves the borrowing.  

\subsection{Defining a core language for relational symbolic execution}

Database programs can be written in many different programming languages, using different DBMS interfaces, and taking advantage of the numerous features of the various SQL dialects. This strong variability makes it necessary to define a framework to properly study our relational symbolic execution approach. Practically, we have chosen to define a formal subset of the language made of Java  empowered by JDBC \cite{jdbc} and ISO SQL \cite{SQL}, which are all very popular technologies. The defined sublanguage contains base SQL primitives as well as a Turing-complete subset of Java. The symbolic execution of many of the Java constructs not considered here has been studied elsewhere (see e.g. \cite{cbt4,Java-tool2B}) and is a problem orthogonal to this work, which studies the \emph{integration} of SQL handling into classic symbolic execution. The considered sublanguage allows integers as the only datatype for the  values stored in the database. Non-linear integer arithmetic operations are not permitted by the sublanguage. The sole impact of allowing other datatypes, like bit-vectors, reals, timestamps or strings, as well as of allowing non-linear arithmetics, would be that our algorithm would produce bit-vector, real, timestamp, string and non-linear arithmetics constraints, in addition to the linear integer constraints produced here. Solving more heterogeneous constraint sets is evaluated in the discussion part of the paper.  

Concretely, our language describes OLTP business use cases. Such an use case is composed of the manipulated database's DDL schema and of the code of the Java method implementing the use case. The database schema describes a set of tables composed of integer attributes. Each table has a primary key and the attributes can be constrained by foreign key constraints and check constraints. The code of the method can contain if-then-else blocks, while loops and return statements. It can contain integer and list (local) variable assignments, with typical operators for lists and linear operators for integers. Assertions can be declared in the code. The method interacts with the database through SQL base statements whose structure is entirely described statically, and through SQL base primitives for transaction management. Statements writing in the database can throw runtime exceptions if the write operation violates the DDL schema. Such exceptions can be caught. The method receives as input parameters a JDBC connection to the database, a set of integer lists and an input scanner for integers. The lists model any structured group of inputs transmitted to the code at method call. The scanner models the method's access to  simple data from the "outside world", like user prompt, network access, etc.

Figure \ref{sample} provides a sample use case written in the language considered here. This sample describes a database with two tables: one for library shelves and one for the books stored in each of these shelves. The total number of books stored in a shelf is saved for each shelf. The Java method adds a set of new books to the database and updates the shelves' books counts. If a book is added to a non-existent shelf, then the shelf is itself added to the database as well. The books are inserted one by one in isolated transactions. If a transaction was successful, the code of the added book is saved in a list, which is returned at the end of the method's execution.

\subsection{A formal definition for our core language}
\label{62}

We define now precisely -- using a BNF grammar --what subset of the Java/SQL syntax our algorithm can execute symbolically. This description is made with the implicit requirement that the written code is well-typed according the relevant standard Java and SQL typing rules.

In the next paragraphs, the chosen notation for the BNF is standard but includes some additional meta-symbols: \{...\} (grouping), $^?$ (zero or one times), * (zero or more times) and $^+$ (one or more times). When a single nonterminal appears several times in a single production, subscript notation enables distinguishing between the occurrences. 

\subsubsection{OLTP use case}

An OLTP use case is composed of the SQL DDL code of the manipulated database and of the code of the Java method implementing it.

\begin{grammar}
{\small
<oltp-use-case> ::= <sql-ddl> <Java-method> 
}	
\end{grammar}	  

\subsubsection{Database schema}

The relational database schema is a list of table definitions. This list can be empty, in which case the method is a classical method that works independently of any database. In the list, each table is identified by its name, contains at least one attribute and endorses exactly one primary key. Foreign keys and additional check constraints can be declared for a table. A row in a table cannot be deleted or see its primary key value modified as long as there exists at least another row in the database that references it (ON DELETE/UPDATE NO ACTION). The semantics of all the schema creation primitives conforms to the standard \cite{SQL} SQL DDL.

\begin{grammar}
{\small
<sql-ddl> ::= <table>*

<table> ::= "CREATE TABLE" <id> "("<att>$^+$ <p-key> <f-key>* <chk>*")"";"

<att> ::= <id> "INTEGER" "NOT" "NULL" ","

<p-key> ::= "CONSTRAINT" <id>$_{cst}$ "PRIMARY" "KEY" "(" <id>$_{att}$ ")"

<f-key> ::= ",""CONSTRAINT" <id>$_{cst}$ "FOREIGN" "KEY" "(" <id>$_{att}$ ")" "REFERENCES" <id>$_{tab}$ "(" <id>$_{refid}$ ")"

<chk> ::= ",""CHECK" "("<id> \{"<" | "=" | ">"\} <integer>")"

<id> ::= \{a |...| z | A |...| Z\}\{a |...| z | A |...| Z  | 0 |...| 9\}*

<integer> ::= -$^?$ \{1 | ... | 9\}\{0 | ... | 9\}$^*$ | 0\}	
}	
\end{grammar} 

\subsubsection{Method signature and body}

We consider Java methods manipulating only internal variables and parameters. Variables can only be typed as \lit{int}, \lit{Java.util.List<Java.lang.Integer>} or \lit{Java.sql.ResultSet}. The method receives as input parameters a connection to the database (typed as \lit{Java.sql.Connection}), a scanner (typed as \lit{Java.util.Scanner}) and some lists of integers (typed as \lit{Java.util.List<Java.lang.Integer>}), where two distinct list parameters cannot reference a single list object. Its return type can be either \lit{void}, \lit{int} or \lit{Java.util.List<Java.lang.Integer>}. 

\begin{grammar}
{\small
<Java-method> ::= <type> <id> "("<db-con>","<inp> <parameters>") throws SQLException"
			   \\ " " \hspace{2cm} "{" <stmt>* "}"

<type> ::= "void" \alt "int" \alt "List<Integer\>"

<db-con> ::= "Connection con"

<inp> ::= "Scanner in"

<parameters> ::= \{"," "List<Integer\>" <id> \}*
}	
\end{grammar}	 

The connection with the database is assumed reliable and every SQL statement, being well formed, is processed for its effect on the database. The semantics of all the Java constructs conforms to the classical Java specification and documentation. The semantics of all SQL statements conforms to the standard \cite{SQL} SQL specification.

\subsubsection{Common statements and list management} Common condition, loop, assertion and assignment statements, as well as common integer expressions and boolean conditions can be used. Lists can be manipulated with \lit{add(int)}, \lit{remove(int)}, \lit{get(int)} and \lit{size()} methods. The \lit{Java.util.ArrayList} implementation of these methods is assumed to be used.  A list variable can be \lit{null}.

\begin{grammar}
{\small
<stmt> ::=  "if" "("<cond>")" "{"<stmt>$_{then}$*"} "\{"else {"<stmt>$_{else}$*"}"\}$^?$";" 
		  \alt "while" "("<cond>")" "{" <stmt>* "}"";" 
		  \alt "assert" <cond> ";"
		 \alt \{"int" | "List<Integer>"\}$^?$ <id> "=" <expr>";"
		 \alt <id>".add(" <int-expr> ")"";"
		 \alt <id>".remove(" <int-expr> ")"";"
		 \alt "return" <id>";"
}	
\end{grammar}	  

\begin{landscape}
\begin{figure*}
\centering
\caption{SQL and Java/JDBC code of an use case adding books in a library database.}
\lstset{language=SQL,frame=none,columns=flexible}
\lstset{
 morekeywords={references}
}
\begin{flushright} 
 \begin{minipage}{1.38\textwidth}
{
\begin{tabular}{|cc|} 
\hline
\begin{lstlisting} 
CREATE TABLE shelf (
 id INTEGER NOT NULL,
 numberOfBooks INTEGER NOT NULL,
 CONSTRAINT sPK PRIMARY KEY (id),
 CHECK(numberOfBooks > 0));
\end{lstlisting}
&  
\begin{lstlisting} 
CREATE TABLE book (
 code INTEGER NOT NULL,
 shelfId INTEGER NOT NULL,
 CONSTRAINT bPK PRIMARY KEY (code),
 CONSTRAINT bFK FOREIGN KEY(shelfId) 
                                     REFERENCES shelf (id));                      
\end{lstlisting}
\\\hline
\end{tabular}}
\end{minipage}
\end{flushright} 
\lstset{language=Java,frame=single,columns=flexible,showstringspaces=false,numbers=left}
\begin{flushright} 
 \begin{minipage}{1.38\textwidth}
{
\begin{lstlisting}
List<Integer> addBooks (Connection con, Scanner in, List<Integer> newBooks) throws SQLException {
 int i = 0;  List<Integer> addedBooks = new ArrayList<Integer>();
 while ( !(newBooks == null) & (i < newBooks.size()) ) {
   int error = 0;  int theShelf = in.nextInt();
   ResultSet shelves = con.createStatement().executeQuery("SELECT id FROM shelf WHERE id="+theShelf);
   if ( ! shelves.next() )
    con.createStatement().execute("INSERT INTO shelf VALUES ("+theShelf+",1)");
   else 
    con.createStatement().execute("UPDATE shelf SET numberOfBooks=numberOfBooks+1 WHERE id ="+theShelf);
   try { 
    con.createStatement().execute("INSERT INTO book VALUES ("+newBooks.get(i)+","+theShelf+")");
   } catch (SQLException e) { 
    error = 1; 
   };
   if (error==0) { 
    con.commit(); 
    addedBooks.add(newBooks.get(i)); 
   } else 
    con.rollback(); 
   i = i + 1;
 };  
 return addedBooks; }
\end{lstlisting}}
\end{minipage}
\end{flushright} 
\label{sample}
\end{figure*}
\end{landscape}

\begin{grammar}
{\small 
<cond> ::= "true" 
		\alt "false"
		\alt "(""!" <cond>")"
		\alt "("<cond>$_1$  \{"&" | "|"\} <cond>$_2$")" 
		\alt "("<int-expr>$_1$ \{"<" | "==" | ">"\} <int-expr>$_2$")" 
  		\alt "("<id> "==" "null"")"

<expr> ::= <int-expr> | <list-expr>
		
<int-expr> ::= <id> 
		\alt <integer>  
		\alt  "("<int-expr>$_1$  \{"+" | "-"\} <int-expr>$_2$")" 
		\alt "("<id>".get(" <int-expr> ")"")"
		\alt "("<id>".size()"")"
		
<list-expr> ::= <id> 
		\alt "null" 
		\alt  "new ArrayList<Integer>()"
}	
\end{grammar}	  

\paragraph{Interacting with the outside world} The scanner parameter of the method can be used to get integer data from the "outside world" (user prompt, network access, reading from a file, etc.). This interaction is assumed to always succeed, without any technical problem. We have thus the following new alternative for the \synt{stmt} non-terminal:

\begin{grammar}
{\small		  
<stmt> ::= \{"int"\}$^?$ <id> "= in.nextInt();"
}	
\end{grammar}	  

\paragraph{Reading data from the database} Data can be read from the database using simple SQL queries. The obtained ResultSet can be accessed using the \lit{next()} and \lit{getInt(String)} methods. We have thus the following alternatives for existing terminals:

\begin{grammar}
{\small	
<stmt> ::=  \{" ResultSet "\}$^?$ <id> "=" "con.createStatement().executeQuery(\"" <select-query> "\");" 
		  \alt <id>".next();"	

<int-expr> ::=  <id>$_{tab}$".getInt(\"" <id>$_{att}$ "\")"		  

<cond> ::= "("<id>".next(" <int-expr> ")"")"}
\end{grammar}	  

as well as the following new terminals:

\begin{grammar}
{\small		  
<select-query> ::= "SELECT" \{<id>$_i$","\}*<id>$_n$ "FROM" <id>$_{tab}$ \{ "WHERE" <db-cond> \}$^?$""

<db-cond> ::= "("<db-cond>$_1$  \{"AND" | "OR"\} <db-cond>$_2$")"
		\alt "(""NOT" <db-cond>")"
		\alt "("<id> \{"<" | "=" | ">"\} <db-int-expr>")"
		
<db-int-expr> ::= <id> 
		\alt <integer>  
		\alt  "("<db-int-expr>$_1$  \{"+" | "-"\} <db-int-expr>$_2$")" 
		\alt "\"+(" <int-expr> ")+\""	}
\end{grammar}

Integer expressions appearing in SQL code (\synt{db-int-expr}) can be parameterised by integer constants computed dynamically by the Java method. For example, in:

{\footnotesize
\lstset{language=Java,frame=none,columns=flexible,showstringspaces=false,numbers=none}
\begin{lstlisting} 
void example (Connection con, Scanner in) throws SQLException {
 int n = 0;  
 ResultSet r= con.createStatement().executeQuery("SELECT id FROM table WHERE id="+(n)+"");
 r.next();
 int p = r.getInt("id");
}
\end{lstlisting}
}

The charachter string that will be effectively sent to and processed by DBMS is:

{\footnotesize
\lstset{language=Java,frame=none,columns=flexible,showstringspaces=false,numbers=none}
\begin{lstlisting} 
SELECT id FROM table WHERE id=0
\end{lstlisting}
}

The parametric constants can depend on the method's inputs, like in:

 {\footnotesize
\lstset{language=Java,frame=none,columns=flexible,showstringspaces=false,numbers=none}
\begin{lstlisting} 
void example (Connection con, Scanner in) throws SQLException {
 int n = in.nextInt();  
 ResultSet r= con.createStatement().executeQuery("SELECT id FROM table WHERE id="+(n)+"");
}
\end{lstlisting}
}

\paragraph{Writing data into the database}  Data can be written into the database using simple SQL INSERT, UPDATE or DELETE statements. If the execution of such a statement provokes a violation of one of the database schema integrity constraints, the database remains unmodified by the statement, an exception is thrown within the method and its execution is stopped. Such exceptions should be caught using a try/catch structure.

\begin{grammar}
{\small
<stmt> ::= 	 "con.createStatement().execute(\"" <db-write> "\");" 
		 \alt "try { con.createStatement().execute(\"" <db-write> "\"); }"  
		     \\   "catch (SQLException e)" 
		     \\   "{" <stmt>* "};"
		 
<db-write> ::= "INSERT" "INTO" <id> "VALUES" "(" \{ <int-expr>$_i$"," \}* <int-expr>$_n$ ")" 
		     \alt "UPDATE" <id>$_{tab}$ "SET" <id>$_{att}$ "=" <db-int-expr> \{ "WHERE" <db-cond> \}$^?$
		     \alt "DELETE" "FROM" <id> \{ "WHERE" <db-cond> \}$^?$
}	
\end{grammar}	  

\paragraph{Transaction management} SQL transactions are managed through the classical commit and rollback statements. We suppose that a new transaction is automatically started at the beginning of the method's execution. The first call to commit or rollback will end this transaction and then starts a new one. Any subsequent call to commit or rollback will end the current transaction and start a new one. When a commit statement is executed, it makes permanent all the changes made to the database by the method since the current transaction was started. When a rollback statement is executed, it restores the database to its state at the start of the current transaction. We suppose that all the changes made to the database since the last transaction was started are automatically committed at the end of the method's execution.

\begin{grammar}
{\small	    
<stmt> ::= 	  "con.commit""("")" ";" \alt "con.rollback""("")" ";"  
}	
\end{grammar}	  

\subsection{Relational symbolic execution algorithm}
\label{algorithm}

\subsubsection{Inputs and outputs}

The proposed algorithm, which is described in this section, receives as inputs the SQL DDL code describing the schema of the database, the Java code of the method under test and an execution path through this method. It produces as output a constraint system mixing classical constraints with relational constraints. Solutions to this system are such that when the method is executed with respect to any of these solutions, its execution will follow the given path. 

Coupling this algorithm with any existing technique (e.g. \cite{paths,cbt1,concolic,dse1}) able to explore a set of paths to test in the method's control flow graph will enable generating test data for these paths. The set of paths for which test data are computed, as well as the process used to explore these paths, are thus parameters of the method that we propose. This enables the method to be used within the context of different code coverage criteria \cite{adequacy-zhu}. \\

The execution path received as input by our algorithm is supposed to be a finite path in the method's control flow graph \cite{adequacy-zhu}. It defines which branches were taken at each of the encountered if statements, how many times the body of each encountered loop was executed (this number must be finite), which assertions were violated and, for each encountered try/catch statement, whether the catch clause was executed. 

Our algorithm translates the path into a constraint system, combining the \emph{path constraint} with the \emph{database schema integrity constraints}, expressed in SMT-Lib \cite{smt2}, a widely adopted language used as the standard language for many SMT solvers. The generated constraints fit into the SMT-Lib AUFLIA logic, i.e. they can involve quantified array, uninterpreted functions and linear integer arithmetic constraints.  

Solving the constraint system generated by the algorithm for a complete path enables finding values for both the inputs and outputs of the analysed Java method. The inputs include the content of each database table at the start of the method's execution, the value of every list received as argument by the method, and a value for the part of the input stream that is consumed during the method's execution. The outputs include the content of each database table at the end of the method's execution, the final value of each of the argument lists of the method, and possibly the value returned by the return statement. If the constraint system produced for a given path has no solution, this means that the path is infeasible. As the produced constraints are written in a quantified logic that is not generally decidable \cite{smt-alloy}, it can happen that for a given path the solver may neither be able to find a solution for the generated constraint system, nor be able to establish that such a solution does not exist. This is coherent with the problem being not computable in general.

\subsubsection{Algorithm principle}

The algorithm performs a symbolic execution of the path received as input. Each of the successive values taken by the method's variables and by the database tables during the execution of the path is represented by corresponding symbols and defined by constraints. 

First, symbolic execution generates constraints over the symbols representing the initial values of the database tables. These constraints state that, initially, each table contains data that conform to the database schema integrity constraints. 

Then, symbolic execution analyses one by one the method's statements in the order specified by the path. Each time a statement sets or changes the value of a method variable or database table, symbolic execution generates constraints over the symbols representing the new value. These constraints define how the new value can be computed from the values of the database tables and method variables before the statement's execution. Moreover, every time the value of a database table is changed, constraints are also added to state that the new value satisfies the database schema integrity constraints.  

Finally, every time an if, while, assert or try/catch statement is encountered, symbolic execution generates an additional constraint over the symbols such that when the method is executed with respect to values satisfying this constraint, the execution is guaranteed to take the considered path.  

\subsubsection{Execution example}

In the following paragraphs, we illustrate the execution of the algorithm over the sample use case given in Figure~\ref{sample} (page \pageref{sample}). We detail each step of the symbolic execution process over the path where the while loop is executed once, the else branch of both the if statements is taken, and the catch clause of the try/catch is executed (lines 1-6, 8-15, 18-21, 3 and 22). 
 
At each step, we present the rules used by our algorithm to generate the corresponding SMT-Lib symbols and constraints. It should be noted that SMT-Lib syntax is inspired by the S-expressions from Lisp, where classical expressions like $2 + 2 + 2$ are written as $(+~2~(+~2~2))$.
All the rules that are part of the complete set of rules defining our symbolic execution algorithm for SimpleDB are presented during one of these steps, and/or described  formally in a set of tables available at the end of this section. \\

The first step executed by our algorithm is to generate SMT-Lib symbols and constraints for the SQL DDL code of the database schema. For the database schema described in Figure~\ref{sample}, the generated SMT-Lib code is detailed in the frame below. For the reader's convenience, the corresponding Java/SQL code will be reminded as a preliminary comment in the SMT-Lib code. 

First of all, the algorithm generates new symbol types for the kind of objects stored in each table defined by the schema (the lines prefixed by (0) in the SMT-Lib code below). It will then generate symbols and constraints describing the input content of each of these tables. The used modelling is inspired by the one proposed in \cite{smt-alloy} for relational types. First, a symbol is created (1) to represent the initial set of objects in each table. Typed as a boolean function, it returns true for each object present in the input content of the table. Symbols typed as integer functions are then generated (2) to associate to each object in the table one of its attribute values. Finally, constraints are generated to enforce on this input content all the check constraints (3), primary key constraints (4), and foreign key constraints (5) defined in the schema. 

Note that the original SQL table and attribute names, as well as the original Java variable names, are used as SMT-Lib symbols, suffixed by the natural number 1 (e.g. $book1$ or $id1$ in the SMT-Lib code below), which represents the fact that the current symbols represent the initial values of the represented tables, attributes or variables. Subsequent values of a same table, attribute or variable will be represented by the same symbol suffixed with successive numbers. Moreover, as detecting corner cases linked to arithmetic overflow is not a priority of our work, we have used SMT-Lib symbols typed as mathematical integers to represent efficiently the fixed-width integers used the code.

{\small
\begin{lstlisting}[language=smtlib,frame=single,columns=flexible,escapechar=\%]
; %CREATE TABLE book (%
; % code INTEGER NOT NULL,%
; % shelfId INTEGER NOT NULL,%
; % CONSTRAINT bPK PRIMARY KEY (code),%
; % CONSTRAINT bFK FOREIGN KEY(shelfId) REFERENCES shelf (id));%                            

; %CREATE TABLE shelf (%
; % id INTEGER NOT NULL,%
; % numberOfBooks INTEGER NOT NULL,%
; % CONSTRAINT sPK PRIMARY KEY (id),%
; % CHECK(numberOfBooks > 0));% 
; New types for tables
(0) (declare-sort book)
(0) (declare-sort shelf)
; Input content of table Book
(1) (declare-fun book1 (book) Bool)
(2) (declare-fun shelfId1 (book) Int)
(2) (declare-fun code1 (book) Int)
(4) (assert (forall ((a book) (b book)) 
	(=> (and (and (book1 a) (book1 b)) (= (code1 a) (code1 b))) (= a b))))

; Input content of table Shelf
(1) (declare-fun shelf1 (shelf) Bool)
(2) (declare-fun numberOfBooks1 (shelf) Int)
(2) (declare-fun id1 (shelf) Int)
(3) (assert (forall ((a shelf)) (> (numberOfBooks1 a) 0)))
(4) (assert (forall ((a shelf) (b shelf)) 
	(=> (and (and (shelf1 a) (shelf1 b)) (= (id1 a) (id1 b))) (= a b))))

; Foreign keys
(5) (assert (forall ((a book)) 
	(=> (book1 a) (exists ((b shelf)) (and (shelf1 b) (= (shelfId1 a) (id1 b)))))))
\end{lstlisting}}

The second step executed by our algorithm is to define a new SMT-Lib symbol type (called BoundedList) for lists of integers. All the symbols that will be subsequently generated to represent the value of a variable typed as a Java list will be part of this new SMT-Lib type. A BoundedList symbol represents a record composed of three fields: the $isNull$ field is typed as boolean, the $size$ field is typed as integer and the $elements$ field is typed as array of integers. If the $isNull$ field is true, then the symbol represents the Java \lit*{null} value. Otherwise, the field $size$ represents the size of the list, and the field $elements$ represents an array whose indexes $0$ to  $(size-1)$ contain the elements of the list in the right order. 

{\small
\begin{lstlisting}[language=smtlib,frame=single,columns=flexible,escapechar=\%]
(declare-datatypes () 
   ((BoundedList (mk-bounded-list (isNull Bool) (size Int) (elements (Array Int Int))))))
\end{lstlisting}
}

It should be noted that the BoundedList symbol type is defined using an algebraic datatype declaration, where "mk-bounded-list" will be the ad-hoc constructor for list objects. If datatype declarations are handled by several SMT solvers, the related syntax has not been standardised in SMT-Lib. In this work, we use the datatype notation available in the Microsoft Z3 SMT solver \cite{z3}. \\

The third step executed by our algorithm is to define symbols (typed as BoundedList) for the initial content of each list parameter of the method. For the example method considered in this section, the following code is generated:

{\small
\begin{lstlisting}[language=smtlib,frame=single,columns=flexible,escapechar=\%]
; % INPUT PARAMETER: List<Integer> newBooks %
(declare-const newbooks1  BoundedList)
(assert (=> (not (isNull newbooks1)) (>= (size newbooks1) 0)))
\end{lstlisting}
}

It should be noted that the second constraint enforces a general property of BoundedList objects, described earlier. However, it will be enforced one by one for each BoundedList object. This is an optimisation compared to using a more general constraint, quantified over the set of all possible BoundedList objects. In order to solve such a constraint, the solver would indeed be required to instantiate correctly the quantifier by itself, making the constraint more costly to solve.\\

The algorithm can then proceed with the symbolic execution of the method. It follows the path received as input and considers all the statements one by one. In the case of our example, the two first statements to be executed are assignments. Symbolic execution for assignment creates a new symbol of the correct type to represent the new value of the assigned variable (1) and generates constraints to specify that this new symbol contains the value computed by evaluating the expression on the right of the \lit{=} symbol (2). In the particular case where a list variable is assigned to a different list variable, the shared content of the two variables is represented by a single symbol. 

{\small
\begin{lstlisting}[language=smtlib,frame=single,columns=flexible,escapechar=\%]
; %int i = 0;%
(1) (declare-const i1 Int)
(2) (assert (= i1 0))
\end{lstlisting}}{\small
\begin{lstlisting}[language=smtlib,frame=single,columns=flexible,escapechar=\%]
; %List<Integer> addedBooks = new ArrayList<Integer>();%
(1) (declare-const addedbooks1 BoundedList)
(2) (assert (not (isNull addedbooks1)))
(2) (assert (= (size addedbooks1) 0))
\end{lstlisting}}

The next statement in the path is a while statement. As the path specifies that the loop body must be executed, a constraint is generated to specify that the loop condition at this point of time should be true: 

{\small
\begin{lstlisting}[language=smtlib,frame=single,columns=flexible,escapechar=\%]
; %ENTERING LOOP: while ( !(newBooks == null) \& (i < newBooks.size()) ) {%
(assert (and (not (isNull newbooks1)) (< i1 (size newbooks1))))
\end{lstlisting}}

Then the algorithm proceeds with symbolic execution of the statements in the loop body, as specified within the input path. The first statement is an assignment statement:

{\small
\begin{lstlisting}[language=smtlib,frame=single,columns=flexible,escapechar=\%]
; %int error = 0;  %
(declare-const error1 Int)
(assert (= error1 0))
\end{lstlisting}}

Symbolic execution for use of the input scanner simply creates a new symbol to represent the scanned value:

{\small
\begin{lstlisting}[language=smtlib,frame=single,columns=flexible,escapechar=\%]
; %int theShelf = in.nextInt(); %
(declare-const theshelf1 Int)
\end{lstlisting}}

Symbolic execution for select statements creates new symbols to represent the content of the ResultSet variable. A first symbol (1) describes the number of rows returned by the select query. These rows are available through a second symbol (2) which is a function that returns them in the order in which they are returned by the ResultSet: $(shelves1List~0)$ will be the first returned row, $(shelves1List~1)$ the second one and so on.

{\small
\begin{lstlisting}[language=smtlib,frame=single,columns=flexible,escapechar=\%]
; % ResultSet shelves %
(1) (declare-const shelves1Size Int)
(2) (declare-fun shelves1List (Int) shelf)
\end{lstlisting}}

Constraints are then generated to specify that a row is part of the ResultSet if and only if it is part of the current content of the table on which the select query is executed and that it enforces the WHERE condition of the select query. In order to do so, the modelling proposed in \cite{smt-alloy} for constraining the content and cardinality of relations, in a way so that the constraints can be effectively solved by Z3, is used. Following \cite{smt-alloy}, new constraints are added (1) to define a function $shelves1InvertedList$ which is the inverse of $shelves1List$. This function is used (1) to ensure that $shelves1List$ defines a one to one correspondence between the integers $0\leqslant i \leqslant shelves1Size$ and the elements in the ResultSet. Helper code (2) is added to ensure an efficient pattern-based quantifier instantiation \cite{ematchingz3} by the solver, using the \lit*{:pattern} keyword \cite{smtlib2}.  

{\small
\begin{lstlisting}[language=smtlib,frame=single,columns=flexible,escapechar=\%]
; % = con.createStatement().executeQuery("SELECT id FROM shelf WHERE id="+theShelf); %
(1) (declare-fun shelves1InvertedList (shelf) Int)
(2) (declare-fun shelves1Trigger (Int) Bool)
(1) (assert (and (>= shelves1Size 0)  
   	      (=> (= shelves1Size 0) 
		   (forall ((c shelf)) (not (and (shelf1 c) (= (id1 c) theshelf1)))))))
(1) (assert (forall ((c shelf)) 
   (=> (and (shelf1 c) (= (id1 c) theshelf1)) 
   	(and (>= (shelves1InvertedList c) 0) (< (shelves1InvertedList c) shelves1Size)))))
(1) (assert (forall ((c shelf)) 
   (=> (and (shelf1 c) (= (id1 c) theshelf1)) 
        (= c (shelves1List (shelves1InvertedList c))))))
(1) (assert (forall ((i Int)) 
   (=> (and (>= i 0) (< i shelves1Size)) 
   	(= i (shelves1InvertedList (shelves1List i))))))
(1) (assert (forall ((i Int)) 
   (! (=> (and (>= i 0) (< i shelves1Size)) 
   	   (and (shelf1 (shelves1List i)) (= (id1 (shelves1List i)) theshelf1))) 
(2)   :pattern (shelves1Trigger i))))
(2) (assert (=> (>= 0 shelves1Size) (shelves1Trigger 1)))
(2) (assert (forall ((i Int)) 
   (! (=> (and (>= i 0) (< i shelves1Size)) 
    	   (shelves1Trigger (+ i 1))) 
   :pattern (shelves1Trigger i))))
\end{lstlisting}}

As the path specifies that the else branch of the if statement must be executed this time, a constraint is generated to specify that the condition of the if should be false, i.e. that \lit*{shelves.next()} should return true. 

For each ResultSet object, the algorithm records the number of times the \lit*{next()} method has been called on this object. This value represents the index increased by one of the row pointed by the cursor of the ResultSet at the current execution state of the path. When the boolean value returned by the \lit{next()} method is used in an if or while condition, this value states if the number of rows in the ResultSet is greater or equal to the number of times the \lit{next()} method has been called so far on this ResultSet. In this case, \lit*{shelves.next()} will return true if the ResultSet \lit*{shelves} contains at least one row (as \lit*{shelves.next()} has been called once on the ResultSet):

{\small
\begin{lstlisting}[language=smtlib,frame=single,columns=flexible,escapechar=\%]
; %TAKING ELSE BRANCH OF: if ( ! shelves.next() )%
(assert (>= shelves1Size 1))
\end{lstlisting}
}

Symbolic execution for update creates a new symbol (1) typed as an integer function, that will replace the previous symbol associating the attribute value to each object in the table. As this new symbol is the second one to represent the value of the attribute $numberOfBooks$, it is named $numberOfBooks2$.
A couple of constraints (2)(3) is then generated to relate the old and new attribute values in the table: one for the rows that do not match the WHERE condition (2), and one for those that do (3).   Finally, constraints are added to specify that no integrity constraint was violated during the update. In this case, a  constraint (4) is added to state that the updated attribute values still enforce the check constraint defined in the database schema. 

{\small \begin{lstlisting}[language=smtlib,frame=single,columns=flexible,escapechar=\%]
; %con.createStatement().execute(%
; %                "UPDATE shelf SET numberOfBooks=numberOfBooks+1 WHERE id ="+theShelf);%
(1) (declare-fun numberOfBooks2 (shelf) Int)
(2) (assert (forall ((p shelf)) 
	(=> (or (and (shelf1 p) (not (= (id1 p) theshelf1))) (not (shelf1 p)))  
	     (= (numberOfBooks2 p) (numberOfBooks1 p)))))
(3) (assert (forall ((p shelf)) 
	(=> (and (shelf1 p) (= (id1 p) theshelf1))   
	     (= (numberOfBooks2 p) (+ (numberOfBooks1 p) 1)))))
(4) (assert (forall ((a shelf)) (> (numberOfBooks2 a) 0)))
\end{lstlisting}}

Subsequently, as the path specifies that the catch block of the try/catch statement must be executed, a constraint (1) is added to ensure that the method variables and the database are in a state where the INSERT execution will violate a schema integrity constraint. In this case, the constraint states that the inserted row has a similar primary key as the primary key of an existing row in the table or that the inserted row has a foreign key value that does not reference any existing row in the shelf table. Constraints are also automatically added to ensure that the \lit{get(int)} (2) method does not cause any runtime error.

{\small
\begin{lstlisting}[language=smtlib,frame=single,columns=flexible,escapechar=\%]
; %TAKING THE CATCH BRANCH OF: % 
; % try \{ con.createStatement().execute(  %
; %           "INSERT INTO book VALUES ("+newBooks.get(i)+","+theShelf+")"); %
; %\} catch (SQLException e) \{ %
(1) (assert (or (exists ((a book)) (and (book1 a) 
			       	     (= (code1 a) (select (elements newbooks1) i1)))) 
	    (forall ((a shelf)) (=> (shelf1 a) 
		 		     (not (= (id1 a) theshelf1))))))
(2) (assert (not (isNull newbooks1)))
(2) (assert (>= i1 0))
(2) (assert (< i1 (size newbooks1)))
\end{lstlisting}}

The content of the catch block is then symbolically executed:

{\small
\begin{lstlisting}[language=smtlib,frame=single,columns=flexible,escapechar=\%]
; %error = 1;%
(declare-const error2 Int)
(assert (= error2 1))
\end{lstlisting}}

As the path specifies that the else branch of the if statement must be executed this time, a constraint is generated to specify that the condition of the if should be false:

{\small
\begin{lstlisting}[language=smtlib,frame=single,columns=flexible,escapechar=\%]
; %TAKING ELSE BRANCH OF: if (error==0) %
(assert (not (= error2 0)))
\end{lstlisting}}

Symbolic execution for Rollback statements tells the algorithm to represent the current content of each database table using the symbols that were representing the content of the table just before the last start of a new transaction (saved by the algorithm at the beginning of the method execution and after each call to commit or abort). In this case, the database state is restored to its state at the method start, i.e. the algorithm rewinds the counters for the database symbols and symbols $book1$, $code1$, $shelfId1$, $shelf1$, $id1$ and $numberOfBooks1$  represent the content of the database after the \lit{con.rollback()} statement.

The assignment statement is then symbolically executed:

{\small
\begin{lstlisting}[language=smtlib,frame=single,columns=flexible,escapechar=\%]
; %i = i + 1; %
(declare-const i2 Int)
(assert (= i2 (+ i1 1)))
\end{lstlisting}}

As the path specifies that the loop body must not be executed any more, a constraint is generated to specify that, at this point in time, the loop condition should be false:

{\small
\begin{lstlisting}[language=smtlib,frame=single,columns=flexible,escapechar=\%]
; %ESCAPING LOOP: while ( !(newBooks == null) \& (i < newBooks.size()) ) {%
(assert (not (and (not (isNull newbooks1)) (< i2 (size newbooks1)))))
\end{lstlisting}}

As a return statement is met, the symbolic execution can be stopped and the generated SMT-Lib constraint model can be returned.  The Z3 solver can now be asked to find a valuation for the defined symbols satisfying the constraints. As the algorithm records what symbols represent the initial, respectively final, values of a variable or table, the
input and output values of the method (for the considered path) can easily be extracted from the solution to the constraint system.

For our example, the set of 29 constraints was solved by Z3 in 24ms (1.8 GHz Intel Core i5, 8GB Ram) and the test data that were obtained from the solution to the constraint system are summarised in the following tables:

{\footnotesize
\begin{center}
\begin{tabular}{|p{0.10\textwidth}|p{0.14\textwidth}|p{0.15\textwidth}|}
\hline
 \multicolumn{3}{|l|}{\textbf{Inputs}}  \\
 \hline 
\textbf{Name}&\textbf{Symbol(s)}&\textbf{Value}\\
\hline 
TABLE shelf & CONTENT: 

$shelf1$, 

ATTRIBUTES: 

$id1$

$numberOfBooks1$ & ~ 

 \begin{tabular}{|c|c|}\hline id & n.Books \\\hline 6  & 1  \\\hline 12  & 1  \\\hline \end{tabular} \\
\hline 
TABLE book  & CONTENT: 

$book1$ 

ATTRIBUTES: 

$code1$ 

$shelfId1$ & ~ 

\begin{tabular}{|c|c|}\hline code & s.Id \\\hline 4  & 12 \\\hline \end{tabular}    \\
\hline 
newBooks& $newbooks1$ & $\left[4\right]$  \\
\hline 
in.nextInt()  & $theshelf1$ & $\left[6\right]$  \\
\hline 
\end{tabular}
\begin{tabular}{|p{0.10\textwidth}|p{0.14\textwidth}|p{0.15\textwidth}|}
\hline
 \multicolumn{3}{|l|}{\textbf{Outputs}}  \\
 \hline 
\textbf{Name}&\textbf{Symbolic(s)}&\textbf{Value}\\
\hline 
TABLE shelf & CONTENT: 

$shelf1$, 

ATTRIBUTES: 

$id1$

$numberOfBooks1$ & ~ 

 \begin{tabular}{|c|c|}\hline id & n.Books \\\hline 6  & 1  \\\hline 12  & 1  \\\hline \end{tabular} \\
\hline 
TABLE book  & CONTENT: 

$book1$ 

ATTRIBUTES: 

$code1$ 

$shelfId1$ & ~ 

\begin{tabular}{|c|c|}\hline code & s.Id \\\hline 4  & 12 \\\hline \end{tabular}    \\
\hline 
newBooks  & $newbooks1$ & $\left[4\right]$  \\
\hline 
addedBooks & $addedbooks1$ & $\left[\right]$ \\
\hline 
\end{tabular}~\\
\end{center}
}
\newpage

\subsubsection{Constraint Generation Rules}

For sake of completeness, the following tables define the constraint generation rules used by our algorithm in case of an Insert (table \ref{insert}), Update (table \ref{update}), Delete (table \ref{delete}) and Add/Remove (table \ref{ar}) statement. Table \ref{abbrev} explains the abbreviations to be used in the other tables.

In each of these tables, any declaration of a new symbol leverages a generator providing a fresh symbol identifier, i.e. which has still not been used in the SMT-Lib code generated so far. This is denoted by (for a function declaration):

{\small
\begin{lstlisting}[language=smtlib,columns=flexible,escapechar=@]
(declare-fun @\textsf{freshSym}@ Type)
\end{lstlisting}}

and by (for a constant declaration):

{\small
\begin{lstlisting}[language=smtlib,columns=flexible,escapechar=@]
(declare-const @\textsf{freshSym}@ Type)
\end{lstlisting}}

All the subsequent references to \textsf{freshSym} in the table represent this newly declared symbol. The generated fresh symbols are named according to the naming rule detailed along the example given in the previous subsection. \\

Finally, it should be noted that assertions are handled as if statements. For example :

{\small
\begin{lstlisting}[language=Java,columns=flexible,escapechar=\%,morekeywords={assert}]
assert x == 0;
\end{lstlisting}}

is handled as:

{\small
\begin{lstlisting}[language=Java,columns=flexible,escapechar=\%]
if (!(x == 0))
	End of computation with AssertionError
\end{lstlisting}}

Assertions are particularly useful for test data generation as they let the programmer express additional constraints otherwise non-obvious to the solver.

\begin{table*}
\centering
\caption{SMT-Lib Translation Abbreviations List}
\begin{tabular}{|p{0.17\textwidth}|p{0.79\textwidth}|} \hline
\textbf{Abbreviation}& \textbf{Meaning} \\ \hline
$smt2Of(x)$ & Java condition/expression x translated into a corresponding SMT-Lib condition/expression.  \\ \hline
$smt2Of(x,t,r)$ & SQL condition/expression x evaluated for row r in table t translated into a corresponding SMT-Lib condition/expression.  \\ \hline
\textsf{name}(x) & \textbf{if} (x refers to a database table name) \textbf{then} 

	The symbol that represents the current content of table x 

\textbf{else if} (x refers to a database attribute name) 

	The symbol that represents the current values of attribute x 

\textbf{else if} (x refers to a Java variable name) 

	The symbol that represents the current content of the Java variable x  \\ \hline
\textsf{att}$_i$ & Name of the i$_{th}$ attribute in the list of attributes of table \synt{id} \\ \hline
\textsf{pk} & Name of the primary key attribute of table \synt{id}. \\ \hline
\textsf{pk}$^{pos}$ & Position of primary key in the list of attributes of table \synt{id} \\ \hline
\textsf{fk}$_i^{tab}$ & Name of the table referenced by the i$_{th}$ foreign key in the list of foreign keys of table \synt{id} \\ \hline
\textsf{fk}$_i^{pk}$ & Name of the primary key attribute of the table referenced by the i$_{th}$ foreign key in the list of foreign keys of table \synt{id} \\ \hline
\textsf{fk}$_i^{pos}$ & Position of the foreign key attribute, declared by the i$_{th}$ foreign key in the list of table \synt{id}, in the list of attributes of table \synt{id} \\ \hline
\textsf{ifk}$_i^{tab}$ & Name of the table where is declared the i$_{th}$ foreign key referencing table \synt{id} in the whole schema \\ \hline
\textsf{ifk}$_i^{att}$ & Name of the foreign key attribute declared by the i$_{th}$ foreign key referencing table \synt{id} in the schema\\ \hline
\textsf{co}$_i^{pos}$ & Position of the attribute constrained by the i$_{th}$ check constraint declared in table \synt{id}  \\ \hline
\textsf{co}$_i^{right}$ & Inverted right part of the i$_{th}$ check constraint declared in table \synt{id} (i.e. inverted right part of "a > 0" is "< 0") \\ \hline
\end{tabular}
\label{abbrev}
\end{table*}

\lstset{frame=none}

\begin{table*}
\centering
\caption{SMT-Lib constraints generation rules for INSERT statements}
{\small
\begin{tabular}{|p{0.96\textwidth}|} \hline
\lit*{INSERT INTO} \synt{id} \lit*{VALUES} \lit*{(}\synt{int-expr}$_1$ , ... , \synt{int-expr}$_i$ , ... , \synt{int-expr}$_n$\lit*{)} 
\\ \hline \vspace{-4mm} \begin{lstlisting}[language=Java,columns=flexible,escapechar=@]
if (@\textit{no exception thrown in path for this INSERT}@) {
\end{lstlisting}  \vspace{-2mm} \begin{lstlisting}[language=smtlib,columns=flexible,escapechar=@]
; Inserted primary key value does not already exist
(assert(forall((a@$\synt{id}$@))(=>(@$\textsf{name}(\synt{id})$@ a)(not(= (@\textsf{name}(\textsf{pk})@ a) @$smt2Of($\synt{int-expr}$_{\textsf{pk}^{pos}})$@)))))
; Inserted values constrained by the @$i^{th}$@ foreign key reference existing rows
(assert(exists((a @$\textsf{fk}^{tab}_i$@))(and(= (@$\textsf{name}({\textsf{fk}^{pk}_i})$@ a) @$smt2Of($\synt{int-expr}$_{\textsf{fk}^{pos}_i})$@) (@$\textsf{name}({\textsf{fk}^{tab}_i})$@ a))))
; Symbol for new table content
(declare-fun @\textsf{freshSym}@ (@\synt{id}@) Bool)
; Constraints describing new table content
(assert (forall ((a @\synt{id}@)) (=> (@$\textsf{name}$(\synt{id})@ a) (@\textsf{freshSym}@ a))))
(assert (exists ((a @\synt{id}@)) (and (= (@\textsf{att}$_i$@ a) @$smt2Of($\synt{int-expr}$_i)$@) (@\textsf{freshSym}@ a))))
(assert (forall ((a @\synt{id}@)) 
 (=>(and(not (@$\textsf{name}$(\synt{id})@ a))(not (= (@\textsf{att}$_i$@ a) @$smt2Of($\synt{int-expr}$_i)$@)))(not (@\textsf{freshSym}@ a)))))
; No duplicate inserted row 
(assert (forall ((a @\synt{id}@) (b @\synt{id}@)) 
 (=>(and(and (@\textsf{freshSym}@ a) (@\textsf{freshSym}@ b)) (= (@\textsf{pk}@ a) (@\textsf{pk}@ b))) (= a b))))
\end{lstlisting}\vspace{-2mm} \begin{lstlisting}[language=Java,columns=flexible,escapechar=\%]
} else { 
%// \textit{Logical disjunction between every possible constraint} %
%// \textit{violation given the database schema and this insert:}%
\end{lstlisting}  \vspace{-2mm} \begin{lstlisting}[language=smtlib,columns=flexible,escapechar=@]
; The inserted primary key value already exists in the table
(exists ((a @\synt{id}@)) (and (@$\textsf{name}(\synt{id})$@ a) (= (@\textsf{name}(\textsf{pk})@ a) @$smt2Of($\synt{int-expr}$_{\textsf{pk}^{pos}})$@)))
; @$i^{th}$@ inserted foreign key value does not reference a row:
(forall ((a @$\textsf{fk}^{tab}_i$@))(=> (@$\textsf{name}({\textsf{fk}^{tab}_i})$@ a) (not (= (@$\textsf{name}({\textsf{fk}^{pk}_i})$@ a) @$smt2Of($\synt{int-expr}$_{\textsf{fk}^{pos}_i})$@))))
; An inserted attribute violates the @$i^{th}$@ check constraint:
(not (@$\textsf{co}^{right}_i$@ @$smt2Of($\synt{int-expr}$_{\textsf{co}^{pos}_i})$@))

}
\end{lstlisting}\vspace{-4mm}  \\ \hline
\end{tabular}
}
\label{insert}
\end{table*}

\begin{table*}
\centering
\caption{SMT-Lib constraints generation for UPDATE statements}
{\small
\begin{tabular}{|p{0.95\textwidth}|} \hline
\lit*{UPDATE} \synt{id} \lit*{SET} \synt{id}$_{att}$ \lit*{=} \synt{db-int-expr} \lit*{WHERE} \synt{db-cond}
\\ \hline \vspace{-4mm} \begin{lstlisting}[language=Java,columns=flexible,escapechar=\%]
if (%\textit{no exception thrown in path for this UPDATE}%) {
\end{lstlisting}  \vspace{-2mm} \begin{lstlisting}[language=smtlib,columns=flexible,escapechar=@]
; Symbol for new attribute values
(declare-fun @\textsf{freshSym}@ (@\synt{id}@) Int)	
; Constraints describing new attribute values
(assert (forall ((a @\synt{id}@)) (=> (or (and (@$\textsf{name}$(\synt{id})@ a) (not @$smt2Of($\synt{db-cond},\synt{id},a$)$@)) 
	(not (@$\textsf{name}$(\synt{id})@ a))) (= (@\textsf{freshSym}@ a) (@$\textsf{name}(\synt{id}_{att})$@ a)))))
(assert (forall ((a @\synt{id}@)) (=> (and (@$\textsf{name}$(\synt{id})@ a) @$smt2Of($\synt{db-cond},\synt{id},a$)$@)  
 	(= (@\textsf{freshSym}@ a) @$smt2Of($\synt{db-int-expr},\synt{id},a$)$@))))
; Update on attribute constrained by @$i^{th}$@ foreign key not leave pending references
(assert (forall ((a @\synt{id}@)) (=> (@$\textsf{name}(\synt{id})$@ a) 
	(exists ((b @$\textsf{fk}^{tab}_i$@)) (and (@$\textsf{name}({\textsf{fk}^{tab}_i})$@ b) (= (@\textsf{freshSym}@ a) (@$\textsf{name}({\textsf{fk}^{pk}_i}$)@ b)))))))
; Update on attribute constrained by primary key not leaves duplicate attribute values
(assert (forall ((a @\synt{id}@) (b @\synt{id}@)) 
  (=> (and (and(@$\textsf{name}(\synt{id})$@ a) (@$\textsf{name}(\synt{id})$@ b)) (= (@\textsf{freshSym}@ a) (@\textsf{freshSym}@ b))) (= a b))))
; Update on primary key referenced by @$i^{th}$@ foreign key does not leave pending references
(assert (forall ((a @$\textsf{ifk}^{tab}_i$@)) (=> (@$\textsf{name}({\textsf{ifk}^{tab}_i}$)@ a) 
 	(exists ((b @\synt{id}@))(and (@\textsf{name}(\synt{id})@ b) (= (@$\textsf{name}({\textsf{ifk}_i^{att}}$)@ a) (@\textsf{freshSym}@ b)))))))
; Update on attribute constrained by @$i^{th}$@ check constraint does not violate the constraint
(assert (forall ((a @\synt{id}@)) (@$\textsf{co}^{right}_i$@ (@\textsf{freshSym}@ a))))
\end{lstlisting}\vspace{-2mm} \begin{lstlisting}[language=Java,columns=flexible,escapechar=\%]
} else { 
%// \textit{Logical disjunction between every possible constraint} %
%// \textit{violation given the database schema and this update:}%
\end{lstlisting}  \vspace{-2mm} \begin{lstlisting}[language=smtlib,columns=flexible,escapechar=@]
; Update on primary key leads to duplicate attribute values
(exists ((a @\synt{id}@) (b @\synt{id}@)) (and(and (@$\textsf{name}(\synt{id})$@ a) (and (@$\textsf{name}(\synt{id})$@ b) (not (= a b)))) 
	(or (and @$smt2Of($\synt{db-cond},\synt{id},a$)$@ (and @$smt2Of($\synt{db-cond},\synt{id},b$)$@ 
		(= @$smt2Of($\synt{db-int-expr},\synt{id},a$)$@ @$smt2Of($\synt{db-int-expr},\synt{id},b$)$@))) 
	 (and (not @$smt2Of($\synt{db-cond},\synt{id},a$)$@) (and @$smt2Of($\synt{db-cond},\synt{id},b$)$@ 
	      (= (@$\textsf{name}(\synt{id}_{att})$@ a) @$smt2Of($\synt{db-int-expr},\synt{id},b$)$@))))) )
; Update on primary key referenced by the @$i^{th}$@ foreign key leaves pending references
(exists ((a @\synt{id}@) (b @\synt{id}@)) (and (and (@$\textsf{name}(\synt{id})$@ a) (@$\textsf{name}({\textsf{ifk}^{tab}_i})$@ b))
 	(and (and (not (= (@$\textsf{name}(\synt{id}_{att})$@ a) @$smt2Of($\synt{db-int-expr},\synt{id},a$)$)@) 
		(= (@$\textsf{name}(\synt{id}_{att})$@ a) (@$\textsf{name}({\textsf{ifk}^{pk}_i})$@ b))) @$smt2Of($\synt{db-cond},\synt{id},a$)$@)))
; Update on attribute constrained by @$i^{th}$@ foreign key leaves pending references
(exists ((a @\synt{id}@)) (and (and (@$\textsf{name}$(\synt{id})@ a) @$smt2Of($\synt{db-cond},\synt{id},a$)$@)
  (not (exists ((b @$\textsf{name}({\textsf{fk}^{tab}_i})$@)) (= (@$\textsf{name}({\textsf{fk}^{pk}_i})$@ b) @$smt2Of($\synt{db-int-expr},\synt{id},a$)$@))))) 
; Update on attribute constrained by @$i^{th}$@ check constraint violates the constraint
 (exists ((a @\synt{id}@)) (and (and (@$\textsf{name}(\synt{id})$@ a) @$smt2Of($\synt{db-cond},\synt{id},a$)$@) 
		(not (@$\textsf{co}^{right}_i$@ @$smt2Of($\synt{db-int-expr},\synt{id},a$)$@))))
		
}
\end{lstlisting} \vspace{-3mm} \\ \hline
\end{tabular}
}
\label{update}
\end{table*}

\begin{table*}
\centering
\caption{SMT-Lib constraints generation for DELETE statements}
{\small
\begin{tabular}{|p{0.95\textwidth}|} \hline
\lit*{DELETE FROM} \synt{id} \lit*{WHERE} \synt{db-cond}\\ \hline
\vspace{-4mm} \begin{lstlisting}[language=Java,columns=flexible,escapechar=\%]
if (%\textit{no exception thrown in path for this DELETE}%) { 
\end{lstlisting}  \vspace{-2mm} \begin{lstlisting}[language=smtlib,columns=flexible,escapechar=@]
; Symbol for new table content
(declare-fun @\textsf{freshSym}@ (@\synt{id}@) Bool)
; Constraints describing new table content
(assert (forall ((a @\synt{id}@)) 
	(= (@\textsf{freshSym}@ a) (and (@$\textsf{name}(\synt{id})$@ a) (not @$smt2Of($\synt{db-cond},\synt{id},a$)$@)))))	
; Delete does not leave pending references for @$i^{th}$@ foreign key
(assert(forall ((a @$\textsf{fk}^{tab}_i$@) (b @\synt{id}@)) 
	(=> (and (@$\textsf{name}(\synt{id})$@ b) (and (not (@\textsf{freshSym}@ b)) (@$\textsf{name}({\textsf{ifk}^{tab}_i}$)@ a))) 
		(not (= (@$\textsf{name}({\textsf{pk}}$)@ b) (@$\textsf{name}({\textsf{ifk}^{att}_i})$@ a))))))	
\end{lstlisting}\vspace{-2mm} \begin{lstlisting}[language=Java,columns=flexible,escapechar=\%]
} else { 
%// \textit{Logical disjunction between every possible constraint} %
%// \textit{violation given the database schema and this update:}%
\end{lstlisting}  \vspace{-2mm} \begin{lstlisting}[language=smtlib,columns=flexible,escapechar=@]
; Delete leaves pending references for @$i^{th}$@ foreign key
 (exists ((a @$\textsf{fk}^{tab}_i$@) (b @\synt{id}@)) 
	(and (and (and (@$\textsf{name}(\synt{id})$@ b) (@$\textsf{name}({\textsf{ifk}^{tab}_i})$@ a) ) @$smt2Of($\synt{db-cond},\synt{id},b$)$@) 
 		(= (@$\textsf{name}({\textsf{pk}})$@ b) (@$\textsf{name}({\textsf{ifk}^{att}_i})$@ a)))) 
		
}
\end{lstlisting} \vspace{-3mm} \\ \hline
\end{tabular}
}
\label{dbwrite}
\label{delete}
\end{table*}

\begin{table*}
\centering
\caption{SMT-Lib constraints generation rules for {add(int)} and {remove(int)} statements}
{\small
\begin{tabular}{|p{0.95\textwidth}|} \hline
\synt{id}\lit*{.add(} \synt{int-expr} \lit*{);}\\ \hline
\vspace{-4mm} \begin{lstlisting}[language=smtlib,columns=flexible,escapechar=@]
(declare-const @\textsf{freshSym}@ BoundedList)
(assert (not (isNull @$\textsf{name}(\synt{id})$@)))
(assert (not (isNull @\textsf{freshSym}@)))
(assert (= (size @\textsf{freshSym}@) (+ (size @$\textsf{name}(\synt{id})$@) 1)))
(assert (= (elements @\textsf{freshSym}@) 
	(store (elements @$\textsf{name}(\synt{id})$@) (size @$\textsf{name}(\synt{id})$@) @$smt2Of($\synt{int-expr}$)$@)))
\end{lstlisting}\vspace{-2mm}  \\ \hline
\synt{id}\lit*{.remove(} \synt{int-expr} \lit*{);}\\ \hline
\vspace{-4mm} \begin{lstlisting}[language=smtlib,columns=flexible,escapechar=@]
(declare-const @\textsf{freshSym}@ BoundedList)
(assert (not (isNull @$\textsf{name}(\synt{id})$@)))
(assert (not (isNull @\textsf{freshSym}@)))
(assert (>= (size @$\textsf{name}(\synt{id})$@) 1))
(assert (= (size @\textsf{freshSym}@) (- (size "+oldVar+") 1)))
(assert (>= @$smt2Of($\synt{int-expr}$)$@ 0))
(assert (< @$smt2Of($\synt{int-expr}$)$@ (size @$\textsf{name}(\synt{id})$@)))		
(assert (forall ((i Int)) (=> (and (>= i 0) (< i @$smt2Of($\synt{int-expr}$)$@)) 
	(= (select (elements @$\textsf{name}(\synt{id})$@) i) (select (elements @\textsf{freshSym}@) i)))))
(assert (forall ((i Int)) (=> (and (>= i @$smt2Of($\synt{int-expr}$)$@) (< i (size @\textsf{freshSym}@))) 
	(= (select (elements @$\textsf{name}(\synt{id})$@) (+ i 1)) (select (elements @\textsf{freshSym}@) i)))))		

\end{lstlisting}\vspace{-2mm}  \\ \hline
\end{tabular}
}
\label{ar}
\end{table*}

\newpage

\section{Experimental evaluation}
\label{experimentation}

\subsection{Symbolic execution and path exploration}

Originally introduced in \cite{King-1976}, symbolic execution has been used as the core principle of many test data generation techniques. In some of these techniques (see e.g. \cite{survey-tdg} for an overview), symbolic execution is performed for a finite set of finite paths that are \emph{statically} explored in the control-flow graph, i.e. independent of any concrete input values, in accordance with a given coverage criterion \cite{adequacy-zhu}. By solving the constraints associated to each of the paths, one obtains a set of inputs and outputs that satisfy the given coverage criterion.


More recently, test data generation techniques that combine symbolic execution with a \emph{dynamic} path exploration process have also been proposed (see e.g. \cite{cbt1,concolic,dse1}). The point of using a dynamic path exploration process is to provide seamlessly symbolic execution with concrete values to replace the statements for which constraints cannot be generated (like proprietary API calls) or handled by the solver (as they belong to a complex, exotic and/or undecidable logic). This process is called \emph{concretisation} and is deeply discussed and evaluated in \cite{dynvsstat}. In practice, the program is first executed on concrete inputs to produce concrete outputs, but the code is instrumented so that symbolic execution is performed together with this normal run of the program, thereby generating the constraint system corresponding to the concrete execution. By flipping some of the constraints among those generated by this symbolic execution, one may produce constraints whose solution describes new concrete inputs triggering the execution of a different path. The process is then repeated with these new inputs until a set of inputs and outputs has been generated for a set of paths covering a sufficient part of the code, again according to some coverage criterion \cite{adequacy-zhu}.

The principle of relational symbolic execution is orthogonal to the path exploration process, and can be combined with both a static or dynamic approach. However, in order to evaluate the ability and efficiency of our technique, we have built a prototype tool that integrates an implementation of the relational symbolic execution presented in the previous section with a static path exploration process.


\subsection{Experimental framework}


Our tool works as follows. Given a method to test, the tool simply performs a depth-first search of the control-flow graph, considering all the paths that execute the body of each loop in the method at most $K$ times, where $K$ is a parameter of the tool.   Consequently, our prototype satisfies a finitely applicable variant of the common path-coverage criterion \cite{adequacy-zhu}, similar to the loop count-K criterion originally proposed in \cite{loopcount}. 

In parallel to this exploration of the control-flow graph, the tool applies the relational symbolic execution algorithm proposed in section \ref{theo}, and solves the produced constraints using the Z3 solver. Input and output data (including database content) are extracted from the solution to the produced path constraints and recorded as constituting test-cases. Once all the paths have been explored, a so-called \emph{test-suite} comprising all the generated test-cases is returned to the user. The tool also keeps a separate list of the paths for which the solver heuristics fail solving the constraints. 

From a technical standpoint, the tool was coded in Java 1.6 and run on a dual core Intel Core i5 processor at 1.8GHz (256 KB L2 cache per core and 3 MB L3 cache) with 8GB of dual channel DDR3 memory at 1600 MHz. The runtime environment was the Oracle JVM 1.6.0_45 under a 32-bit edition of Microsoft Windows 8.1. The version 4.3.0 of the Microsoft Z3 solver was used. \\

The sample code used for our experimental evaluation is composed of eighteen Java methods, performing SQL operations over relational databases. These methods can be divided into four groups:
\begin{enumerate}
\item[\textbullet] The first group contains nine methods. Each of them was crafted to systematically evaluate the correct symbolic execution of one of the different Java and SQL constructs handled by our algorithm. As such, the methods in this group exercise the different behaviours of the integer and list operators, conditional and loop statements, SQL statements and transaction management primitives. 
\item[\textbullet] The second group contains three methods crafted as typical implementations of OLTP use cases. The first method in this group performs repeated manipulations of integers and lists to compare and save some data. The second method performs many interleaved reads and writes in a database containing four tables, representing regular or prospect customers that make purchases of products. The third method mixes SQL statements with traditional Java code and uses SQL transactions. The manipulated database contains two tables that represent authors writing theatre plays.  
\item[\textbullet] The third and fourth groups contain Java methods extracted respectively from UnixUsage\footnote{\url{http://sourceforge.net/projects/se549unixusage}} and RiskIt\footnote{\url{https://riskitinsurance.svn.sourceforge.net}}, the two open-source software that have been used in \cite{xie-vn}, as a basis for evaluating the proposed test generation approach, based on SQL normalisation into native code. UnixUsage is a monitoring application for Unix, manipulating a database with eight tables and thirty one attributes. RiskIt is an insurance application, manipulating a database with thirteen tables and fifty-seven attributes. 
\end{enumerate}

Together, the methods from our testbed constitute a set of five hundred lines of code, containing notably eighty SQL statements (including SELECT, INSERT, UPDATE, DELETE statements, as well as transaction management code), over databases containing up to thirteen tables (subject to primary key, foreign key and check constraints). 

Detailed statistics for each of these methods can be found in table \ref{statsample}, including the value of K used in our tool to limit the loop exploration depth for each method. Given this value of K, the number of paths to be explored is provided for each method, detailing the number of feasible and infeasible ones. The high number of infeasible paths is principally due to the methods in the second group (and, to a lesser extent, the methods for testing conditional constructs from the first group), methods that were particularly constructed in this way for assessing the soundness of our test generation tool. 

Finally, the code of these methods, as well as the generated test data, can be found on the web\footnote{\url{https://staff.info.unamur.be/mmr/scp/}}. 

\subsection{Evaluating the Alloy and Z3 solvers for relational symbolic execution}
\label{82}

\subsubsection{Performance comparison over common samples}

Our first objective with the experiments was to compare the ability of the Alloy and Z3 solvers to solve effectively and efficiently the constraints produced by relational symbolic execution. In order to do so, we used the Alloy version of our tool from \cite{moi-scam13} and the SMT-Lib version presented here to generate test cases for the methods of the first and second groups from our testbed.\\   

The Alloy solver basically solves the constraints by setting upper bounds on the scope of the possible integers, as well as on the cardinality of the possible different rows which can appear in the solution for each different table. This makes finite the set of possible solutions, which can then be exhaustively explored, by transforming the constraints into an equisatisfiable SAT instance, solved by a dedicated SAT solver. As a consequence, if no solution is found with the default minimal bound value, our tool should repeat the process recursively with a higher value as bound, until a time-out is reached. When the time-out is reached, the tool considers the underlying path as unfeasible. The time-out value should thus be long enough to enable finding a proper model for feasible paths. At the same time, it should not be too long to detect infeasible paths in minimal time. Finding an optimal time-out value appeared to be very difficult in practice, as it depends on the size and complexity of the constraints to solve.

In the left part of table \ref{resultsSDB}, we synthesise the results obtained with the Alloy version of the tool, over the feasible paths of the methods from groups one and two. For each method, the table provides the minimal bound values enabling Alloy to find inputs for all the feasible paths in the method. It also provides the total number of constraints generated for these feasible paths and the minimal time within which the solver was able to solve these constraints. The methods named "Integers", "Update", "Delete" and "Clients and Products" involve either large integer values or repeated actions on a single table. As a consequence, the constraints generated for these methods require large enough bound values to be solved, which had in practice to be found manually, using a costly trial and error approach. Moreover, the results show that once the bounds are increased, the size of the SAT instance generated by the Alloy solver quickly explodes, while the solving time blows up from dozens of milliseconds to several minutes. Finding inputs for the "Handling Data" and the "Play Catalog" methods, with their thousands of paths, revealed to be intractable in reasonable time, with the Alloy solver and without any manual help.\\

In accordance with \cite{smt-alloy}, where Z3 is used as a complement to the Alloy solver for proving instances unsatisfiable, the Z3 solver was able to detect almost instantaneously the unsatisfiability of the generated constraints. As a consequence, the SMT-Lib version of the tool was able to detect properly and efficiently the thousands of unfeasible paths in the methods from groups one and two. As we used a more recent version of Z3 than in \cite{smt-alloy}, involving notably a new model-based quantifier instantiation technique for solving quantified constraints, we could hope that the solver would also be able to find efficiently satisfying models for the generated constraints and thus produce inputs for satisfiable paths. The obtained results are detailed in the right part of table \ref{resultsSDB}, both for feasible and

\begin{landscape}
\begin{table*}
\centering
\caption{Statistics for the selected samples}
\begin{tabular}{|c|c|c|c|c|c|c|} \hline
\textbf{Name} & \textbf{LOCs} & \textbf{Number of} & \textbf{Number}  & \textbf{K Depth} & \multicolumn{2}{|c|}{\textbf{Number of Paths}} \\ \cline{6-7}
& & \textbf{SQL statements} & \textbf{of tables}  & & \textbf{Feasible} & \textbf{Infeasible}  \\ \hline
\multicolumn{7}{|l|}{\textbf{\textit{CRAFTED SAMPLES}}}   \\ \hline
\multicolumn{7}{|l|}{\textbf{\textit{- Language Unit Testing}}}   \\ \hline
Integers & 7 & 0 & 0 & 3 & 1 & 0 \\ \hline
Lists & 9 & 0 & 0 & 3 & 1 & 0 \\ \hline
If and While & 30 & 0 & 0 & 1 & 1 & 319\\ \hline
Conditions & 30 & 0 & 0 & 1 & 1 & 319 \\ \hline
Select & 23 & 7 & 2 & 3 & 1 & 63 \\ \hline
Insert & 14 & 7 & 2 & 3 & 1 & 7 \\ \hline
Update & 20 & 11 & 2 & 3 & 1 & 15 \\ \hline
Delete & 6 & 3 & 2 & 3 & 1 & 1  \\ \hline
Transactions & 13 & 2 & 2  & 3 & 2 & 6 \\ \hline
\multicolumn{7}{|l|}{\textbf{\textit{- Realistic OLTP Methods}}}   \\ \hline
Handling Data & 45 & 1 & 1 & 5 & 23 & 3513 \\ \hline
Clients and Products & 56 & 18 & 4 & 5 & 4 & 32\\ \hline
Play Catalog & 72 & 4 & 2 & 2 & 63 & 5569 \\ \hline 
\multicolumn{7}{|l|}{\textbf{\textit{SAMPLES EXTRACTED FROM OPEN-SOURCE SOFTWARE}}}   \\ \hline
\multicolumn{7}{|l|}{\textbf{\textit{- UnixUsage}}}   \\ \hline
courseNameExists & 7 & 1 & 8 & 1 & 2 & 0 \\ \hline
getCommandsByCategory & 10  & 1 & 8  & 1 & 2 & 0 \\ \hline
getCourseIDByName & 10  & 1 & 8 &  & 2 & 0\\ \hline
getDepartmentIDByName & 11 & 1 & 8 & 1 & 2 & 0  \\ \hline 
\multicolumn{7}{|l|}{\textbf{\textit{- RiskIt}}}   \\ \hline
createNewUser & 91 & 7 & 13 & 2 & 3 & 21\\ \hline
deleteUsers & 55 & 16 & 13 & 2 & 2 & 181 \\ \hline \hline
\textbf{ALL METHODS} & 509 & 80 & up to 13 & up to 5 & 113 & 10046 \\ \hline
\end{tabular}
\label{statsample}
\end{table*}
\end{landscape}

\noindent infeasible paths. The Z3 solver was able to find inputs for each feasible path in milliseconds, always outperforming the Alloy solver, even when minimal bound values were used in the Alloy version. As an example, the Alloy version took more than thirty eight minutes and a half to find test inputs for the four feasible paths of method "Clients and Products", when the Z3 version was able to do it in three seconds and a half. Finding inputs for the "Handling Data" and the "Play Catalog" methods, which was intractable using Alloy, became possible in less than two minutes using Z3.

\subsubsection{Practical scalability limits}

Our second objective with the experiments was to estimate the practical scalability limits of current solver technology for the generated constraints. In order to do so, we have measured the number of generated constraints and the constraint solving time for relational symbolic execution (using Z3) of the following method, containing a single execution path, traversing a linear block of SQL statements. These statements have been selected to provide a balanced mix of typical queries and DML statements. DML statements act on fields subject to integrity constraints.

{\small
\begin{lstlisting}[language=SQL,frame=single,columns=flexible,showstringspaces=false,morekeywords={REFERENCES}]
CREATE TABLE t1 (
idt1 INTEGER NOT NULL,
fieldt1 INTEGER NOT NULL,
CONSTRAINT t1PK PRIMARY KEY (idt1),
CHECK(idt1 > 0));
CREATE TABLE t2 (
idt2 INTEGER NOT NULL,
fieldt2 INTEGER NOT NULL,
CONSTRAINT t2PK PRIMARY KEY (idt2),
CONSTRAINT t2FK FOREIGN KEY (fieldt2) REFERENCES t1(idt1),
CHECK(idt2 > 0));
\end{lstlisting}

\begin{lstlisting}[language=Java,frame=single,columns=flexible,showstringspaces=false]
void test (Connection con,Scanner in) throws SQLException {
int i = 1;
con.createStatement().execute("INSERT INTO t1 VALUES ("+i+","+i+")");
con.createStatement().execute("INSERT INTO t1 VALUES ("+(i+1)+","+(i+1)+")");
con.createStatement().execute("INSERT INTO t2 VALUES ("+i+","+i+")");
con.createStatement().execute("INSERT INTO t2 VALUES ("+(i+1)+","+(i+1)+")");
int input1 = in.nextInt();
ResultSet result1 = con.createStatement().executeQuery("SELECT idt1 
							FROM t1 
							WHERE fieldt1="+i);
result1.next();
con.createStatement().execute("DELETE FROM t2 WHERE idt2="+input1);
con.createStatement().execute("UPDATE t2 
				SET fieldt2 = 1+fieldt2 
				WHERE idt2 < "+(2+result1.getInt("idt1")));
input1 = in.nextInt();
con.createStatement().execute("DELETE FROM t1 WHERE idt1="+input1); }
\end{lstlisting}}

\begin{figure}[h]
\begin{center}
\includegraphics[width=0.65\textwidth]{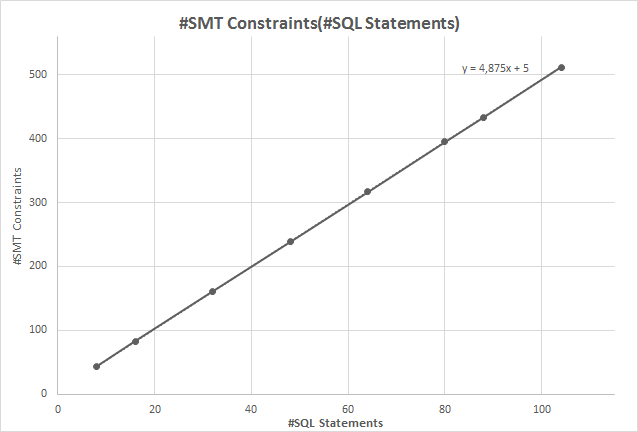}
\caption{Constraint number as a function of the number of SQL statements in the path}
\label{constraintsfig}
\end{center}
\end{figure}

\begin{figure}[h]
\begin{center}
\includegraphics[width=0.70\textwidth]{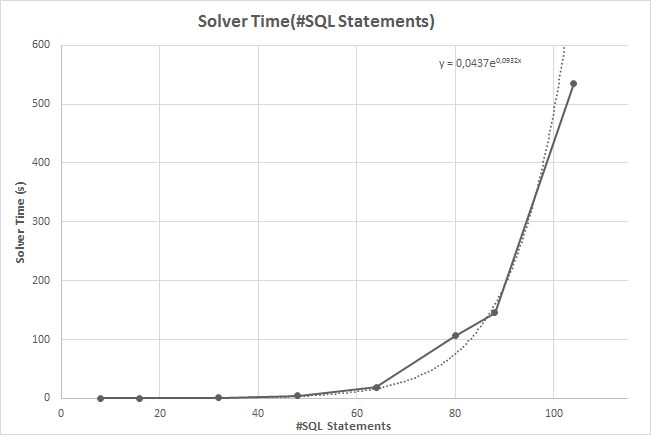}
\caption{Solving time as a function of the number of SQL statements in the path}
\label{timefig}
\end{center}
\end{figure}

\noindent Then, the measurement was repeated for a similar path, but where the block of SQL statements was executed twice in turn: a first time normally and a second time with a value of i increased by 2. The statements of the second round were executed directly on the database resulting from the first round so that they can modify the rows it contains as well.  The process was then repeated with four rounds and so on. The obtained measurements are reported in figure \ref{constraintsfig} and \ref{timefig}. These graphs show the number of constraints and the constraint solving time, as a function of the number of SQL statements executed in the path. 
The number of constraints evolves linearly with the number of SQL statements. The constraint solving time increases much more quickly with the number of SQL statements. The solving time starts to rise around 60 statements and becomes important above 100 statements, on our Intel Core i5 at 1.8GHz.

\begin{landscape}
\begin{table*}
{\small
\centering
\caption{Z3 and Alloy statistics for the methods of the first and second groups}
\begin{tabular}{|c||c|c|c||c|c|c|c|} \hline
\textbf{Name} & \multicolumn{3}{c||}{\textbf{{Feasible Paths to Alloy}}}    & \multicolumn{4}{c|}{\textbf{{All Paths to SMT-Lib}}} \\ \cline{2-8}
& \textbf{Cardinality} & \textbf{Cons-} & \textbf{Solver}  & \multicolumn{2}{c|}{\textbf{{Feasible Paths}}}  & \multicolumn{2}{c|}{\textbf{{Unfeasible Paths}}}    \\ \cline{5-8}
& \textbf{Bound} & \textbf{-traints} & \textbf{Time}  & \textbf{Constraints} & \textbf{Solver Time} &\textbf{Constraints} & \textbf{Solver Time}\\ \hline
\multicolumn{8}{|l|}{\textbf{\textit{CRAFTED SAMPLES}}}   \\ \hline
\multicolumn{8}{|l|}{\textbf{\textit{- Language Unit Testing}}}   \\ \hline
Integers & 1 but 5 int & 6 & 129ms & 4 & 16ms & 0 & 0ms  \\ \hline
Lists & 4 & 10 & 186ms  & 45 & 16ms & 0 & 0ms  \\ \hline
If and While & 1 & 17 & 46ms & 15 & 15ms & 4241 & 3s 989ms  \\ \hline
Conditions & 1 & 19 & 37ms & 19 & 16ms & 5325 & 4s 345ms  \\ \hline
Select & 3 & 41 & 221ms & 61 & 47ms & 4131 & 2s 355ms  \\ \hline
Insert & 1 & 23 & 40ms &  22 & 16ms & 214 & 124ms  \\ \hline
Update & 7 but 8 int & 39 & 6m 6s 299ms & 41 & 63ms & 703 & 765ms   \\ \hline
Delete & 1 but 8 int & 16 & 51s 638ms & 15 & 15ms & 17 & 16ms  \\ \hline
Transactions & 1 & 38 & 137ms  & 44  & 30ms & 148 & 93ms  \\ \hline
\multicolumn{8}{|l|}{\textbf{\textit{- Realistic OLTP Methods}}}   \\ \hline
Handling Data & - & - & - & 2154 & 641ms & 487120 & 1m 3s 504ms   \\ \hline
Clients and Products &  8 but 16 clients & 502  & 38m 37s 506ms & 612 & 3s 470ms & 4068 & 7s 587ms   \\ \hline
Play Catalog & - & - & - & 4343 & 1s 361ms & 527625 & 1m 35s 67ms \\ \hline
\end{tabular}
\label{resultsSDB}}
\end{table*}
\end{landscape} 

\subsection{Benchmarking relational symbolic execution against normalisation}
\label{83}

Our last objective with the experiments was to compare the performance of relational symbolic execution with the approach proposed in \cite{xie-vn}. In this last approach, the tested code is preprocessed to translate the SQL code into native program code, before applying a classical symbolic execution process. In order to perform this comparison, we have selected four simple Java methods from the testbed used by \cite{xie-vn} and used our tool on them. These methods are those of the third group from our testbed. 

Our tool was asked to generate tests with the same statement coverage as offered by \cite{xie-vn}. In order to propose a fair comparison, the total generation time for each method was measured on a Pentium 4 configuration (3.06GHz, 1GB RAM, Microsoft Windows XP 32-bit), similar (but admittedly not identical) to the one used in \cite{xie-vn}. The obtained results are synthesised in the first part of table \ref{resultsCompa}. Results for the normalisation-based tool are extracted from \cite{xie-vn}. While the results are of course not fully conclusive as both tests were run on different (but comparable) machines, we can nevertheless observe our tool to be between one and two orders of magnitude faster than \cite{xie-vn}, generating in each case tests in a few seconds compared to more than one minute and a half.

A first reason for such a performance difference between both tools is that \cite{xie-vn} normalises the SQL code into native code before trying to generate tests. This requires time and strongly increases the number of paths to be explored, as each SQL statement is translated into native code that introduces new branches and cycles in the control-flow graph. Secondly, \cite{xie-vn}  is built upon the Microsoft Pex tool for constraint-based testing, which makes use of a dynamic path exploration, requiring to effectively run the code for each generated test case. Our research prototype tool makes use of a static path exploration and does not need to run the code, which also has an impact on our performance comparison. However, this impact is limited by fact that the methods are very small, involving between 7 and 11 simple statements, with only one simple SQL (SELECT) statement, one loop and no conditional per method. Finally, it should be noted that our tool handles the string attributes appearing in the methods as integer ones. This is possible as the tested methods do not use string-specific operators.

\subsection{Additional experiments over open-source code}
\label{84}

UnixUsage and RiskIt provide a useful source of open-source code for evaluating our testing tool. As a consequence, we have also extracted some methods involving not only SELECT statements, as in \cite{xie-vn}, but also SQL statements writing into the database, in order to assess our tool. These methods are those of the fourth group from out testbed. 

Test data generation for these methods required extending our symbolic execution algorithm to handle some new parts of Java and SQL, used in  RiskIt, or to simulate them the currently handled sublanguage. In particular, the management of tables with no primary key or with multiple-attribute primary key was integrated in the algorithm and string management operations were simulated using either integers or lists of integers. As the second part of table~\ref{resultsCompa} shows, correct test data were generated in just a few seconds and, moreover, manual reviewing of the automatically generated test data enabled detecting a possible fault in the code of RiskIt, where a runtime error is thrown when the method \lit*{createNewUser} is called on inputs where the inserted job does not reference any existing occupation or industry.

\begin{landscape}
\begin{table*}
{\small
\centering
\caption{Analysis time for the methods of the third and fourth groups}
\begin{tabular}{|c|c|c|c|c|} \hline
\textbf{Name} & \multicolumn{2}{|c|}{\textbf{Analysis via SynDB \cite{xie-vn}}} & \multicolumn{2}{|c|}{\textbf{Analysis via Relational Symbolic Execution}} \\ \cline{2-5}
& \textbf{Covered Statements} & \textbf{Analysis Time} & \textbf{Covered Statements}  &\textbf{Analysis Time} \\ \hline
\multicolumn{5}{|l|}{\textbf{\textit{SAMPLES EXTRACTED FROM OPEN-SOURCE SOFTWARE}}}   \\ \hline
\multicolumn{5}{|l|}{\textbf{\textit{- UnixUsage}}}   \\ \hline
courseNameExists & 7 & 1m 41s & 7  & 4s 984ms \\ \hline
getCommandsByCategory & 10  & 1m 30s &  10 & 1s 454ms \\ \hline
getCourseIDByName & 10  & 1m 33s & 10 & 2s 891ms  \\ \hline
getDepartmentIDByName & 11 & 1m 43s & 11  & 1s 578ms  \\ \hline 
\multicolumn{5}{|l|}{\textbf{\textit{- RiskIt}}}   \\ \hline
createNewUser & - & - & 91 & 11s 294ms \\ \hline
deleteUsers & - & - & 55 & 5s 39ms  \\ \hline
\end{tabular}
\label{resultsCompa}}
\end{table*}
\end{landscape}

\section{Discussion}
\label{conclusion}

\subsection{Synthesis of research contributions}

%
In this work, we proposed an approach for enabling the direct symbolic execution of SQL code into constraints. This is a non-trivial extension to traditional symbolic execution because of the complex structure of relational databases and the complex behaviour of SQL statements. Given a database program mixing traditional code with SQL statements, each database table manipulated by the program is modelled as a variable typed as a relation and each SQL statement as a relational operation over both these relational variables and the traditional variables of the program. A classical symbolic execution process can then be applied to produce sets of mixed relational and classical constraints over symbols representing the values of both the classical and relational variables of the program. The resulting path constraints can be unified with the data integrity constraints from the database schema. Any solution to the resulting constraint system for a path describes input and output data for the program, including a valid database content, with respect to which the program can be executed and is guaranteed to follow the same path for which the constraints were generated. 

A symbolic execution algorithm based on this principle has been completely detailed for a precise subset of Java and SQL. This language enables writing Java methods that use SQL statements and transactions to read and write in a relational database; the latter typically subject to data integrity constraints. The algorithm has been designed to test rather small methods, acting on a limited number of tuples in the database, as they are typically written in programs acting in an OLTP context, to implement single business use cases. Given the schema of the database, the code of the method and an execution path in this method, the algorithm performs the symbolic execution of the path and produces the corresponding constraints in the standard SMT-Lib language.

The algorithm has been used in a prototype tool to generate test data for a number of methods, including some open-source code. The generated constraints have been solved using the Microsoft Z3 solver. The experiments show that the technique is able to generate test data for all the considered methods, seamlessly and in reasonable time. In particular, the results dramatically improve the scope of the approach, compared to the strategy based on the Alloy solver, proposed in our previous work \cite{moi-1,moi-scam13}. These results provide both an experimental confirmation and new elements to the research presented in \cite{smt-alloy}, where Z3 was used to prove the unsatisfiability of Alloy constraints. Our experiments can indeed be seen as some kind of a case study for \cite{smt-alloy}, which confirms the conclusions of \cite{smt-alloy} about proving unsatisfiable instances. Moreover, our results show that versions of Z3 more recent than the one used in \cite{smt-alloy}, including new model-based quantifier instantiation techniques, can outperform the Alloy solver also in model finding for satisfiable instances. Finally, our measurements showed that the approach may face scalability issues outside the context of methods implementing OLTP use cases, if it is used as such over pieces of code whose typical execution scenarios involve the processing of a large number of SQL statements.
\newpage

\subsection{Related work}

An early work that has considered test data generation for programs interacting with a relational database is \cite{3}. The paper suggests to transform the program by inserting new classical variables representing the data\-base structure, and translating all SQL statements into native program code acting on these variables. A translation to C++ is provided for some relational operators and for some other SQL mechanisms, like row sorting. Classical white-box testing approaches can then be applied to the normalised program. A conceptually similar but entirely automated technique is proposed in \cite{xie-vn}, where the off-the-shelf Microsoft Pex tool, based on symbolic execution and a dynamic path exploration process, is applied to the normalised version of database programs written in C\#. This latter approach is validated over 39 Java methods (translated to C\# using an automated compiler) extracted from UnixUsage and RiskIt.  These samples involve 32 LOCs per method on average, with a maximum of 108 LOCs, and each method mixes Java statements with one or a few SELECT SQL statements.

Conceptually, normalising SQL code into native code and then applying classical symbolic execution on the result is an alternate approach to ours, where the SQL code is directly compiled into relational constraints during symbolic execution. Replacing a single SQL statement by a piece of native code, simulating its execution by a DBMS, is time-consuming and may strongly increase the number of paths to be explored, compared to the original code \cite{3}. In contrast, in relational symbolic execution, the code must not be preprocessed and the execution paths to be considered are limited to paths in the \emph{original} program code. Our experimental benchmarks, comparing the tool from \cite{xie-vn}, based on normalisation, with our tool, tend to indicate that a direct translation of SQL into constraints provokes a strong performance improvement. Moreover, even if the symbolic execution of INSERT, UPDATE and DELETE statements is conceptually possible using normalisation, \cite{xie-vn} only validates their approach over code containing SELECT statements. Relational symbolic execution has on the contrary be experimented in the presence of SQL code writing into the database, as well as in the presence of SQL transactions. 

On the other hand, the tool proposed in \cite{xie-vn} provides important features that are not handled by ours. First, it deals with more complex SELECT queries. Secondly, it relies on Microsoft Pex, as an off-the-shelf back-end constraint-based testing tool. First, Pex handles natively a large part of the C\# language, where our tool is restricted to a small part of Java. Secondly, Pex provides some support for character strings, which constitute an important datatype in SQL, not handled by our tool. Thirdly, Pex uses a dynamic path exploration process (coupled with heuristics for a smarter covering of large control-flow graphs), making concrete values for the program variables available if necessary. This notably gives a direct access to the concrete SQL code that can be, in some programs, crafted dynamically as a character string, to be parsed and processed by the DBMS. As a consequence, the tool from \cite{xie-vn} can account, at least to some extent, for such dynamically crafted SQL statements, where our tool only considers SQL statements whose structure is completely defined statically. \\

In \cite{13}, the JCUTE symbolic execution tool for Java is provided with the ability to generate database inputs in the presence of simple SELECT queries in the tested code. The principle is to consider the result of any query as an independent array-like input for the program. The size of each of these input arrays, as well as the content of each of their cells can then be accessed in the normal code. As a consequence, the path constraint generated in JCUTE contains constraints over these sizes and contents. Additional constraints are added to the path constraint, enforcing that each row in an input array makes the condition of the WHERE clause from its corresponding SELECT true. Constraints are solved using an ad-hoc solver for strings. Experimentation is discussed for a 16 lines of code database program written in Java involving one query extracted from open-source software.

This approach exhibits some conceptual limitations compared to relational symbolic execution. 
It indeed provides no mean to handle INSERT, UPDATE and DELETE statements, neither to account for transaction management primitives, which are crucial components of database programs. Moreover, \cite{13} only provides a conceptual intuition and no proper mechanism for enforcing the data integrity constraints defined in the database schema. \cite{xie-vn} advocates that such a lack leads to the possible generation of invalid test data and to a poor code coverage. 

Nevertheless, the tool detailed in \cite{13} has also important features that are not present in our tool. First, and contrary to our tool, it deals with null values and, at least partly, with character strings. Secondly, it relies on the JCute constraint-based testing tool. JCute handles a large part of the Java language, including multi-threading, and uses a dynamic path exploration process. As with Pex, this makes the tool from \cite{13} able to benefit from concretisation as well as to account, at least to some extent, for dynamically crafted SQL statements. \\

On a related but complementary level, a substantial amount of work (e.g.  \cite{veryveryold,oldsoa,8,three,otherApproach,6,4,agendabis,12,10,qex,11,coverscalable,qaGen,9}) has been done on how to generate test database content exhibiting some desirable properties, given only the database schema and possibly some queries to be executed over the database. The main difference between our work and these approaches is that they essentially work without considering the control flow of the programs manipulating the database. 

Microsoft Qex \cite{10,qex} is probably the one of these techniques which is the closest to our work, as both approaches are based on the translation of the SQL semantics into SMT-Lib constraints, solved using Z3. However, the two approaches also exhibit important differences. Firstly, Qex only considers input generation for a single SQL SELECT query in isolation, but the handled queries can be more complex than in our tool. Secondly, the two techniques translate the SQL semantics into constraints differently: our tool use predicates and quantifiers to represent relations, while Qex use fixed-size lists of tuples. Qex involves thus a mechanism similar to Alloy, where the solving process must be repeated on increasing size for the input relations, in order to find a solution. Such an approach does not enable proving the constraints unsatisfiable. 

Some database programs are developed to work with data already stored in an existing database. Some papers \cite{15,concolic-first} study the particular situation where classical test inputs must be generated for code manipulating in parallel an existing database with a known input content.

Other work \cite{sqlmutation,21,mutation2} considers \emph{mutation testing} of database programs, where our approach performs \textit{structural testing}. In mutation testing, the quality of the test data is no more measured in terms of code coverage, as in structural testing, but in terms of program fault detecting ability (see \cite{adequacy-zhu} for a discussion). Some works have also focused over testing of non-functional aspects of database programs, like security testing \cite{mutationSecurity}.

\subsection{Threats to validity}

\subsubsection{Internal validity}

The performance comparison between our relational symbolic execution tool and the SQL normalisation tool from \cite{xie-vn} revealed a very strong advantage for relational symbolic execution, over similar hardware configurations. However, the comparison was performed on a limited number of small samples and considering tools with different architectures. We think that the impact of the last element is minor, given the small size of the tested methods, and cannot explain, on its own, how important the measured performance gap is. However, this element, as well as the lack of a comparison over a large number of various complex pieces of code, threaten our ability to conclude, from our experimental work, that a direct translation of SQL into relational constraints intrinsically improves the testing time, compared to compiling SQL into native code before generating constraints.

\subsubsection{External validity}

The tool we have developed handles a limited subset of the Java/SQL syntax and was developed in the context of transactional business use cases, i.e. small pieces of code acting on a limited number of rows in the database. Several elements threaten the generalisability of our approach to handle any piece of code involving SQL. 

Conceptually, our approach can be integrated into any existing symbolic execution tool, to provide this tool with the ability to handle SQL code. An integration into a state-of-art tool, able to handle not all, but a wide range of programs in a mainstream programming language, would strongly improve the practical scope of our technique. However, such an integration needs to be evaluated in practice. 

The handling of a larger part of SQL may require the generation of constraints which could be hard to manage by current solver technology. Moreover, as SQL is a large and complex language, fragmented into several dialects, developing an universal SQL symbolic execution engine, able to handle any piece of SQL code, will be a very hard task. 

Our tool provides no automated way to handle SQL code dynamically crafted by the program, which can be frequent in practice. Alternate tools have shown that we could exploit the concretisation mechanism from a state-of-art symbolic execution tool to alleviate the problem, thanks to the availability of the concrete values for the dynamically crafted SQL code. However, such an approach fails handling cases where the syntactic structure of the crafted SQL depends on the inputs of the tested code.     

Finally, current solver technologies have shown to face a scalability issue when required to solve the constraints generated by our tool for a sufficiently large number of SQL statements. Whether this problem can be solved or at least alleviated, by optimising the constraint generation or by new solving capabilities, remains an open research question.
\newpage

\subsection{Future work}

For further work, we have identified the following three main research directions: 

\begin{itemize}
\item Our tool has demonstrated how a classical symbolic execution mechanism for a typical programming language could be empowered with the ability to generate constraints in the presence of SQL code. As this integration provided promising results, it should now be repeated with a state-of-art tool based on symbolic execution. Microsoft Pex \cite{cbt2} is a particularly appropriate candidate for integration with our work, as it is based as well on the Z3 solver. In addition, such an integration could enable a deeper comparison with the tool from \cite{xie-vn}, which is also based on Pex. The Pex tool is not open-source but provides an extension interface. To the best of our knowledge, available open-source constraint-based testing tools are CREST \cite{heuristicPaths} and KLEE \cite{heuristicPaths2} (for C code) as well as Symbolic Path Finder \cite{Java-tool2B}  (for Java byte-code).  
\item By integrating our approach in a tool like Pex, the resulting tool would of course keep its advanced features, like a large syntax coverage for a modern programming language, as well as a dynamic path exploration process coupled with heuristics to handle large number of paths. In presence of SQL code built dynamically as a character string in the tested code, the use of a dynamic path exploration would make the concrete values of the assembled string elements directly available. This runtime information should be used to recover the complete structure of the executed SQL code, making us able to translate it into constraints, as described in this paper. 

However, such an approach will fail if the program's inputs are used as parts of the syntactic structure of the dynamically-crafted SQL code. Nevertheless, by choosing appropriate concrete values for those parts of the inputs that are used to define the syntactic structure of the dynamic SQL statements, one could produce representative specialised versions of the original program that could be properly evaluated symbolically. Interleaving symbolic execution with such a partial evaluation \cite{partialevalbook} has already been studied in another context by \cite{partialeval}. Detecting which parts of the programs's inputs should be made concrete could benefit from existing work (e.g. \cite{24,25}) over detection of SQL injections attacks. 
\item  In the perspective of testing programs involving many complex SQL statements, a tighter integration with constraint solving techniques would be beneficial to offer a better scalability and a larger scope to the approach. The constraint generator should be tailored so to generate constraints optimised for the internal algorithms of the solver. Conversely, the development of solving algorithms or heuristics tailored to the kind of constraints produced by the symbolic execution of large pieces of complex SQL code should also be considered. 

In particular, SQL enables various operations to be performed over data belonging to various datatypes, such as strings, binary objects, numeric values and timestamps. Symbolic execution of such operations will produce complex constraints over such datatypes. If modern SMT solvers like Z3 can already handle many of these constraints, work should be done to locate the common parts of SQL which will put current solvers into trouble and research should be performed to possibly build a workaround. Solver development is a particularly dynamic research domain.  Notably, research is ongoing (e.g. \cite{z3-string,stringli}) towards a proper solving of string contraints inter-related with other kinds of constraints, in the context of symbolic execution. Similarly, recent works (see e.g. \cite{arithmsolv}) have considered (partial) solving of non-linear integer arithmetic constraints. Other works have also studied the particular problem of multi-granularity temporal constraint solving (see e.g. \cite{csp-temporal}).

However, building optimal constraint generation rules for the whole of SQL is made difficult by the fact that the syntax and semantics of SQL is large and complex, and can vary strongly in practice between different DBMS's versions and manufacturers. A research direction for overcoming these difficulties would be to use a relational algebra as an intermediate language for constraint generation. The symbolic execution engine would compile the original SQL code into a minimal relational algebra, and then the algebraic operators would be translated into logical constraints, using a minimal set of translation rules optimised for the solver. Algorithms translating SQL statements into equivalent combinations of a core set of relational algebra's operators are well-known, in the context of query processing in DBMS design \cite{connollyDB}. In practice, this idea should be refined, as SQL allows non-relational constructs like rows ordering and aggregation, null values etc. The intermediate language should thus be extended by a minimal set of operators for describing common and tractable non-relational parts of SQL. Concretisation could be the last-ditch solution to handle exotic or too complex parts of SQL.
\end{itemize}

These research directions sketch a path towards a broader evaluation of the approach over many various real pieces of code, in an industrial context. \\

Programs to be tested which manipulate an existing database are common in practice. Whether and how our technique could select test data from an existing database, instead of generating them from scratch is an interesting matter of further research.

Constraints typically admit many different solutions. However, our tool uses the arbitrary solution returned as first by the solver. A possible improvement could be to use an optimal solution according to some criterion, like for example, a minimal number of rows in the database. In order to do so, it has recently been announced\footnote{see "Objective Functions in Z3", Nikolaj Bjorner, keynote at CSTVA 2014, Hyderabad, India.}  that Z3 could be provided with the ability to return the solution which optimises a given objective function. Evaluating and integrating this mechanism with our approach is a topic for future research.

Finally, being somehow parametrised with respect to the paths that should be considered, our approach can be used with respect to any traditional code coverage criterion based on the notion of execution path \cite{adequacy-zhu}. Nevertheless, several works \cite{18,22,19,20,21} propose coverage criteria particularly tailored towards testing of database programs. Integrating such criteria into our constraint-based approach is a topic of ongoing research. 

\section*{Acknowledgment}

This work has been funded by the Belgian Fund for Scientific Research (F.R.S.-FNRS). The authors would like to thank P.-Y. Schobbens, V. Englebert and G. Perrouin for useful discussions, C. Aerts for test hardware availability.

\newpage
\section*{Bibliography}

\bibliographystyle{elsarticle-num}
\bibliography{lala}

\begin{thebibliography}{10}
\expandafter\ifx\csname url\endcsname\relax
  \def\url#1{\texttt{#1}}\fi
\expandafter\ifx\csname urlprefix\endcsname\relax\def\urlprefix{URL }\fi
\expandafter\ifx\csname href\endcsname\relax
  \def\href#1#2{#2} \def\path#1{#1}\fi

\bibitem{Jorgensen2008}
P.~C. Jorgensen, Software Testing: A Craftsman's Approach, Third Edition, 3rd
  Edition, Auerbach, 2008.

\bibitem{testing-book-kaner}
C.~Kaner, H.~Q. Nguyen, J.~L. Falk, Testing Computer Software, 2nd Edition,
  Wiley \& Sons, New York, NY, USA, 1993.

\bibitem{Ramler:2006}
R.~Ramler, K.~Wolfmaier, Economic perspectives in test automation: balancing
  automated and manual testing with opportunity cost, in: Proceedings of
  International Workshop on Automation of Software Test, AST '06, ACM, New
  York, NY, USA, 2006, pp. 85--91.

\bibitem{adequacy-zhu}
H.~Zhu, P.~A.~V. Hall, J.~H.~R. May, Software unit test coverage and adequacy,
  ACM Comput. Surv. 29~(4) (1997) 366--427.

\bibitem{King-1976}
J.~C. King, Symbolic execution and program testing, Commun. ACM 19~(7) (1976)
  385--394.

\bibitem{symbex-review1}
P.~D. Coward, Symbolic execution systems a review, Softw. Eng. J. 3~(6) (1988)
  229--239.

\bibitem{symbex-review2}
C.~S. Pasareanu, W.~Visser, A survey of new trends in symbolic execution for
  software testing and analysis, Int. J. Softw. Tools Technol. Transf. 11~(4)
  (2009) 339--353.

\bibitem{symbex-review3}
E.~J. Schwartz, T.~Avgerinos, D.~Brumley, All you ever wanted to know about
  dynamic taint analysis and forward symbolic execution (but might have been
  afraid to ask), in: Proceedings of the IEEE Symposium on Security and
  Privacy, SP '10, IEEE Computer Society, Washington, DC, USA, 2010, pp.
  317--331.

\bibitem{se-synt}
C.~Cadar, P.~Godefroid, S.~Khurshid, C.~S. Pasareanu, K.~Sen, N.~Tillmann,
  W.~Visser, Symbolic execution for software testing in practice: Preliminary
  assessment, in: Proceedings of the 33rd International Conference on Software
  Engineering, ICSE '11, ACM, New York, NY, USA, 2011, pp. 1066--1071.

\bibitem{sybolic-execution}
C.~Cadar, K.~Sen, Symbolic execution for software testing: three decades later,
  Communication of the ACM 56~(2) (2013) 82--90.

\bibitem{cbt1}
P.~Godefroid, N.~Klarlund, K.~Sen, {DART}: directed automated random testing,
  SIGPLAN Not. 40~(6) (2005) 213--223.

\bibitem{yices}
B.~Dutertre, L.~de~Moura, A fast linear-arithmetic solver for dpll(t),
  Proceedings of the 18th International Conference on Computer Aided
  Verification (2006) 81--94.

\bibitem{Cadar06exe}
C.~Cadar, V.~Ganesh, P.~M. Pawlowski, D.~L. Dill, D.~R. Engler, Exe:
  Automatically generating inputs of death, in: In Proceedings of the 13th ACM
  Conference on Computer and Communications Security (CCS), ACM, 2006.

\bibitem{heuristicPaths2}
C.~Cadar, D.~Dunbar, D.~Engler, Klee: Unassisted and automatic generation of
  high-coverage tests for complex systems programs, Proceedings of the 8th
  USENIX Conference on Operating Systems Design and Implementation (2008)
  209--224.

\bibitem{sage}
P.~Godefroid, M.~Y. Levin, D.~Molnar, Sage: Whitebox fuzzing for security
  testing, Queue 10~(1) (2012) 20:20--20:27.

\bibitem{cbt4}
K.~Sen, G.~Agha, {CUTE} and {jCUTE}: concolic unit testing and explicit path
  model-checking tools, in: Proceedings of the 18th international conference on
  Computer Aided Verification, CAV'06, Springer-Verlag, Berlin, Heidelberg,
  2006, pp. 419--423.

\bibitem{Java-tool2B}
C.~S. P\v{a}s\v{a}reanu, P.~C. Mehlitz, D.~H. Bushnell, K.~Gundy-Burlet,
  M.~Lowry, S.~Person, M.~Pape, Combining unit-level symbolic execution and
  system-level concrete execution for testing {NASA} software, in: Proceedings
  of the International Symposium on Software Testing and Analysis, ISSTA '08,
  ACM, New York, NY, USA, 2008, pp. 15--26.

\bibitem{cbt2}
N.~Tillmann, J.~De~Halleux, Pex: white box test generation for .net, in:
  Proceedings of the 2nd international conference on Tests and proofs, TAP'08,
  Springer-Verlag, Berlin, Heidelberg, 2008, pp. 134--153.

\bibitem{dbs}
S.~Abiteboul, R.~Hull, V.~Vianu, Foundations of Databases, Addison-Wesley,
  1995.

\bibitem{qex}
M.~Veanes, N.~Tillmann, J.~de~Halleux, Qex: Symbolic {SQL} query explorer,
  Proceedings of the 16th International Conference on Logic for Programming,
  Artificial Intelligence, and Reasoning (2010) 425--446.

\bibitem{Date:2003}
C.~Date, An Introduction to Database Systems, 8th Edition, Addison-Wesley
  Longman Pub. Co., Inc., Boston, MA, USA, 2003.

\bibitem{3}
M.~Y. Chan, S.~C. Cheung, Testing database applications with {SQL} semantics,
  in: In Proceedings of the 2nd International Symposium on Cooperative Database
  Systems for Advanced Applications, Springer, 1999, pp. 363--374.

\bibitem{xie-vn}
K.~Pan, X.~Wu, T.~Xie, Guided test generation for database applications via
  synthesized database interactions, ACM Transactions on Software Engineering
  and Methodology ?

\bibitem{13}
M.~Emmi, R.~Majumdar, K.~Sen, Dynamic test input generation for database
  applications, in: Proceedings of th International Symposium on Software
  Testing and Analysis, ISSTA, ACM, New York, NY, USA, 2007, pp. 151--162.

\bibitem{moi-1}
M.~Marcozzi, W.~Vanhoof, J.-L. Hainaut, Test input generation for database
  programs using relational constraints, in: Proceedings of the Fifth
  International Workshop on Testing Database Systems, DBTest '12, ACM, New
  York, NY, USA, 2012, pp. 6:1--6:6.

\bibitem{moi-scam13}
M.~Marcozzi, W.~Vanhoof, J.-L. Hainaut, A relational symbolic execution
  algorithm for constraint-based testing of database programs, in: Source Code
  Analysis and Manipulation (SCAM), IEEE 13th International Working Conference
  on, ACM, 2013, pp. 179--188.

\bibitem{alloy-book}
D.~Jackson, Software Abstractions: Logic, Language, and Analysis, The MIT
  Press, 2006.

\bibitem{smt-alloy}
A.~A. El~Ghazi, M.~Taghdiri, Relational reasoning via {SMT} solving, in:
  Proceedings of the 17th international conference on Formal methods, FM'11,
  Springer-Verlag, Berlin, Heidelberg, 2011, pp. 133--148.

\bibitem{smt2}
C.~Barrett, A.~Stump, C.~Tinelli, {The {SMT}-LIB Standard: Version 2.0}, in:
  A.~Gupta, D.~Kroening (Eds.), Proceedings of the 8th International Workshop
  on Satisfiability Modulo Theories (Edinburgh, UK), 2010.

\bibitem{z3}
L.~De~Moura, N.~Bj{\o}rner, {Z3}: An efficient {SMT} solver, in: Proceedings of
  the Theory and Practice of Software, 14th International Conference on Tools
  and Algorithms for the Construction and Analysis of Systems,
  TACAS'08/ETAPS'08, Springer-Verlag, Berlin, Heidelberg, 2008, pp. 337--340.

\bibitem{mbqi}
Y.~Ge, L.~Moura, Complete instantiation for quantified formulas in
  satisfiabiliby modulo theories, in: Proceedings of the 21st International
  Conference on Computer Aided Verification, CAV '09, Springer-Verlag, Berlin,
  Heidelberg, 2009, pp. 306--320.

\bibitem{gray}
J.~Gray, A.~Reuter, Transaction Processing: Concepts and Techniques, 1st
  Edition, Morgan Kaufmann Publishers Inc., San Francisco, CA, USA, 1992.

\bibitem{otherApproach}
C.~Binnig, D.~Kossmann, E.~Lo, Multi-rqp: Generating test databases for the
  functional testing of oltp applications, Proceedings of the 1st International
  Workshop on Testing Database Systems (2008) 5:1--5:6.

\bibitem{jdbc}
Oracle, Jdbc overview:
  \url{http://www.oracle.com/technetwork/java/overview-141217.html}, Website (3
  2014).

\bibitem{SQL}
ISO, Iso/iec 9075 : 2011 information technology, database languages, {SQL}
  (2011).

\bibitem{paths}
S.~Bardin, P.~Herrmann, Pruning the search space in path-based test generation,
  in: Proceedings of the International Conference on Software Testing
  Verification and Validation, ICST '09, IEEE Computer Society, Washington, DC,
  USA, 2009, pp. 240--249.

\bibitem{concolic}
K.~Sen, D.~Marinov, G.~Agha, Cute: a concolic unit testing engine for {C},
  SIGSOFT Softw. Eng. Notes 30~(5) (2005) 263--272.

\bibitem{dse1}
C.~Cadar, D.~Engler, Execution generated test cases: How to make systems code
  crash itself, in: Proceedings of the 12th International Conference on Model
  Checking Software, SPIN'05, Springer-Verlag, Berlin, Heidelberg, 2005, pp.
  2--23.

\bibitem{ematchingz3}
L.~Moura, N.~Bj{\o}rner, Efficient e-matching for {SMT} solvers, in:
  Proceedings of the 21st International Conference on Automated Deduction:
  Automated Deduction, CADE-21, Springer-Verlag, Berlin, Heidelberg, 2007, pp.
  183--198.

\bibitem{smtlib2}
C.~Barrett, A.~Stump, C.~Tinelli, {The {SMT}-LIB Standard: Version 2.0}, in:
  A.~Gupta, D.~Kroening (Eds.), Proceedings of the 8th International Workshop
  on Satisfiability Modulo Theories (Edinburgh, UK), {SMT}-Lib, 2010.

\bibitem{survey-tdg}
J.~Edvardsson, {A survey on automatic test data generation}, in: Proceedings of
  the Second Conference on Computer Science and Engineering in Link\"{o}ping,
  ECSEL, 1999, pp. 21--28.

\bibitem{dynvsstat}
P.~Godefroid, Higher-order test generation, SIGPLAN Not. 46~(6) (2011)
  258--269.

\bibitem{loopcount}
W.~E. Howden, Methodology for the generation of program test data, Computers,
  IEEE Transactions on 100~(5) (1975) 554--560.

\bibitem{veryveryold}
N.~R. Lyons, An automatic data generating system for data base simulation and
  testing, SIGMIS Database 8~(4) (1977) 10--13.

\bibitem{oldsoa}
M.~Robbert, F.~J. Maryanski, Automated test plan generator for database
  applications systems, SIGSMALL/PC Notes 17~(3-4) (1991) 29--35.

\bibitem{8}
C.~Binnig, D.~Kossmann, E.~Lo, Reverse query processing, in: Data Engineering,
  2007. ICDE 2007. IEEE 23rd International Conference on, IEEE Computer Society
  Press, 2007, pp. 506--515.

\bibitem{three}
C.~Binnig, D.~Kossmann, E.~Lo, Testing database applications, in: Proceedings
  of the 2006 ACM SIGMOD International Conference on Management of Data, SIGMOD
  '06, ACM, New York, NY, USA, 2006, pp. 739--741.

\bibitem{6}
J.~Zhang, C.~Xu, S.~C. Cheung, Automatic generation of database instances for
  white-box testing, in: Proceedings of the 25th International Computer
  Software and Applications Conference, COMPSAC '01, IEEE Computer Society,
  Washington, DC, USA, 2001, pp. 161--165.

\bibitem{4}
D.~Chays, Y.~Deng, P.~G. Frankl, S.~Dan, F.~I. Vokolos, E.~J. Weyuker, An
  agenda for testing relational database applications: Research articles,
  Softw. Test. Verif. Reliab. 14~(1) (2004) 17--44.

\bibitem{agendabis}
Y.~Deng, P.~Frankl, D.~Chays, Testing database transactions with agenda, in:
  Proceedings of the 27th International Conference on Software Engineering,
  ICSE '05, ACM, New York, NY, USA, 2005, pp. 78--87.

\bibitem{12}
D.~Willmor, S.~M. Embury, An intensional approach to the specification of test
  cases for database applications, in: Proceedings of the 28th international
  conference on Software engineering, ICSE '06, ACM, New York, NY, USA, 2006,
  pp. 102--111.

\bibitem{10}
M.~Veanes, P.~Grigorenko, P.~Halleux, N.~Tillmann, Symbolic query exploration,
  in: Proceedings of the 11th International Conference on Formal Engineering
  Methods: Formal Methods and Software Engineering, ICFEM '09, Springer-Verlag,
  Berlin, Heidelberg, 2009, pp. 49--68.

\bibitem{11}
C.~de~la Riva, M.~J. Suarez-Cabal, J.~Tuya, Constraint-based test database
  generation for {SQL} queries, in: Proceedings of the 5th Workshop on
  Automation of Software Test, AST '10, ACM, New York, NY, USA, 2010, pp.
  67--74.

\bibitem{coverscalable}
J.~Tuya, M.~J. Suarez-Cabal, C.~de~la Riva, Full predicate coverage for testing
  {SQL} database queries, Softw. Test. Verif. Reliab. 20~(3) (2010) 237--288.

\bibitem{qaGen}
C.~Binnig, D.~Kossmann, E.~Lo, M.~T. {\"O}zsu, Qagen: generating query-aware
  test databases, in: C.~Y. Chan, B.~C. Ooi, A.~Zhou (Eds.), SIGMOD Conference,
  ACM, 2007, pp. 341--352.

\bibitem{9}
S.~A. Khalek, B.~Elkarablieh, Y.~O. Laleye, S.~Khurshid, Query-aware test
  generation using a relational constraint solver, in: Proceedings of the 2008
  23rd IEEE/ACM International Conference on Automated Software Engineering, ASE
  '08, IEEE Computer Society, Washington, DC, USA, 2008, pp. 238--247.

\bibitem{15}
C.~Li, C.~Csallner, Dynamic symbolic database application testing, in:
  Proceedings of the Third International Workshop on Testing Database Systems,
  DBTest '10, ACM, New York, NY, USA, 2010, pp. 7:1--7:6.

\bibitem{concolic-first}
K.~Pan, X.~Wu, T.~Xie, Generating program inputs for database application
  testing, in: Proc. 26th IEEE/ACM International Conference on Automated
  Software Engineering, IEEE Computer Society Press, 2011.

\bibitem{sqlmutation}
J.~Tuya, M.~J. Suarez-Cabal, C.~de~la Riva, {SQL}mutation: A tool to generate
  mutants of {SQL} database queries, in: Proceedings of the Second Workshop on
  Mutation Analysis, MUTATION '06, IEEE Computer Society, Washington, DC, USA,
  2006, pp. 1--.

\bibitem{21}
C.~Zhou, P.~Frankl, Mutation testing for {Java} database applications, in:
  Proceedings of the International Conference on Software Testing Verification
  and Validation, ICST '09, IEEE Comp. Soc., Washington, DC, USA, 2009, pp.
  396--405.

\bibitem{mutation2}
K.~Pan, X.~Wu, T.~Xie, Automatic test generation for mutation testing on
  database applications, in: Automation of Software Test (AST), 8th
  International Workshop on, IEEE Computer Society Press, 2013, pp. 111--117.

\bibitem{mutationSecurity}
D.~Appelt, C.~D. Nguyen, L.~C. Briand, N.~Alshahwan, Automated testing for
  {SQL} injection vulnerabilities: An input mutation approach, in: Proceedings
  of the International Symposium on Software Testing and Analysis, ISSTA 2014,
  ACM, New York, NY, USA, 2014, pp. 259--269.

\bibitem{heuristicPaths}
J.~Burnim, K.~Sen, Heuristics for scalable dynamic test generation, in:
  Proceedings of the 2008 23rd IEEE/ACM International Conference on Automated
  Software Engineering, ASE '08, IEEE Computer Society, Washington, DC, USA,
  2008, pp. 443--446.

\bibitem{partialevalbook}
N.~D. Jones, C.~K. Gomard, P.~Sestoft, Partial Evaluation and Automatic Program
  Generation, Prentice-Hall, Inc., Upper Saddle River, NJ, USA, 1993.

\bibitem{partialeval}
R.~Bubel, R.~H\"{a}hnle, R.~Ji, Interleaving symbolic execution and partial
  evaluation, in: Proceedings of the 8th International Conference on Formal
  Methods for Components and Objects, FMCO'09, Springer-Verlag, Berlin,
  Heidelberg, 2010, pp. 125--146.

\bibitem{24}
W.~G.~J. Halfond, A.~Orso, P.~Manolios, Using positive tainting and
  syntax-aware evaluation to counter {SQL} injection attacks, in: Proceedings
  of the 14th ACM SIGSOFT International Symposium on Foundations of Software
  Engineering, SIGSOFT '06/FSE-14, ACM, New York, NY, USA, 2006, pp. 175--185.

\bibitem{25}
Z.~Su, G.~Wassermann, The essence of command injection attacks in web
  applications, SIGPLAN Not. 41~(1) (2006) 372--382.

\bibitem{z3-string}
Y.~Zheng, X.~Zhang, V.~Ganesh, {Z3}-str: A {Z3}-based string solver for web
  application analysis, in: Proceedings of the 2013 9th Joint Meeting on
  Foundations of Software Engineering, ESEC/FSE 2013, ACM, New York, NY, USA,
  2013, pp. 114--124.

\bibitem{stringli}
T.~Liang, A.~Reynolds, C.~Tinelli, C.~Barrett, M.~Deters, A dpll(t) theory
  solver for a theory of strings and regular expressions, in: A.~Biere,
  R.~Bloem (Eds.), CAV, Vol. 8559 of Lecture Notes in Computer Science,
  Springer, 2014, pp. 646--662.

\bibitem{arithmsolv}
D.~Jovanovic, L.~M. de~Moura, Solving non-linear arithmetic (2012) 339--354.

\bibitem{csp-temporal}
C.~Bettini, X.~Wang, S.~Jajodia, Solving multi-granularity temporal constraint
  networks, Artificial Intelligence 140~(1--2) (2002) 107 -- 152.

\bibitem{connollyDB}
T.~Connolly, C.~Begg, Database Systems: A Practical Approach to Design,
  Implementation, and Management, no. vol.~1 in International computer science
  series, Addison-Wesley, 2005.

\bibitem{18}
W.~G.~J. Halfond, A.~Orso, Command-form coverage for testing database
  applications, in: Proceedings of the 21st IEEE/ACM International Conference
  on Automated Software Engineering, ASE '06, IEEE Computer Society,
  Washington, DC, USA, 2006, pp. 69--80.

\bibitem{22}
G.~M. Kapfhammer, M.~L. Soffa, A family of test adequacy criteria for
  database-driven applications, in: Proceedings of the 9th European software
  engineering conference held jointly with 11th ACM SIGSOFT international
  symposium on Foundations of software engineering, ESEC/FSE-11, ACM, New York,
  NY, USA, 2003, pp. 98--107.

\bibitem{19}
M.~J. Su\'{a}rez-Cabal, J.~Tuya, Using an {SQL} coverage measurement for
  testing database applications, in: Proceedings of the 12th ACM SIGSOFT
  Twelfth International Symposium on Foundations of Software Engineering,
  SIGSOFT '04/FSE-12, ACM, New York, NY, USA, 2004, pp. 253--262.

\bibitem{20}
M.~J. Suarez-Cabal, J.~Tuya, {Structural coverage criteria for testing {SQL}
  queries}, Journal of Universal Computer Science 15~(3) (2009) 584--619.

\end{thebibliography}

\end{document}